\documentclass[twocolumn,showpacs,pre,english,showkeys,10pt,aps]{revtex4-1}
\usepackage{graphicx,amsmath,amssymb}
\usepackage[english]{babel}
\usepackage{hyperref}

\usepackage{graphicx}
\usepackage{epstopdf}

\begin{document}

\title{Efficient numerical methods for the random-field Ising model: \\ Finite-size scaling, reweighting extrapolation, and
computation of response functions}

\author{Nikolaos G. Fytas}

\affiliation{Applied Mathematics Research Centre, Coventry
University, Coventry CV1 5FB, United Kingdom}

\author{V\'{i}ctor Mart\'{i}n-Mayor}

\affiliation{Departamento de F\'{i}sica Te\'{o}rica I, Universidad
Complutense, E-28040 Madrid, Spain}

\affiliation{Instituto de Biocomputaci\'on and
  F\'{\i}sica de Sistemas Complejos (BIFI), E-50009 Zaragoza, Spain}

\begin{abstract}
It was recently shown [Phys. Rev. Lett. {\bf 110}, 227201 (2013)]
that the critical behavior of the random-field Ising model in
three dimensions is ruled by a single universality class. This
conclusion was reached only after a proper taming of the large
scaling corrections of the model by applying a combined approach
of various techniques, coming from the zero- and
positive-temperature toolboxes of statistical physics. In the
present contribution we provide a detailed description of this
combined scheme, explaining in detail the zero-temperature
numerical scheme and developing the generalized
fluctuation-dissipation formula that allowed us to compute
connected and disconnected correlation functions of the model. We
discuss the error evolution of our method and we illustrate the
infinite limit-size extrapolation of several observables within
phenomenological renormalization. We present an extension of the
quotients method that allows us to obtain estimates of the
critical exponent $\alpha$ of the specific heat of the model via
the scaling of the bond energy and we discuss the self-averaging
properties of the system and the algorithmic aspects of the
maximum-flow algorithm used.
\end{abstract}

\pacs{75.10.Nr, 02.60.Pn, 75.50.Lk}

\maketitle

\section{Introduction}
\label{intro}

The random-field Ising model (RFIM) is one of the archetypal
disordered
systems~\cite{imry:75,aharony:76,young:77,fishman:79,parisi:79,cardy:84,imbrie:84,schwartz:85,gofman:93,esser:97,barber:01},
extensively studied due to its theoretical interest, as well as its
close connection to experiments in condensed matter
physics~\cite{belanger:91,rieger:95b,nattermann:98,belanger:98,belanger:83,vink:06}.
In particular, several important systems can be studied through the
RFIM: diluted antiferromagnets in a field~\cite{belanger:98},
colloid-polymer mixtures~\cite{vink:06,annunziata:12}, colossal
magnetoresistance oxides~\cite{dagotto:05,burgy:01}, phase-coexistence
in the presence of quenched
disorder~\cite{cardy:97,fernandez:08,fernandez:12}, non-equilibrium
phenomena such as the Barkhausen noise in magnetic
hysteresis~\cite{sethna:93,perkovic:99} or the design of switchable
magnetic domains~\cite{silevitch:10}, etc.

The existence of an ordered ferromagnetic phase for the RFIM, at
low temperature and weak disorder, followed from the seminal
discussion of Imry and Ma~\cite{imry:75}, when the space dimension
is greater than two ($D
> 2$)~\cite{villain:84,bray:85b,fisher:86b,berker:86,bricmont:87}. This
has provided us with a general qualitative agreement on the sketch
of the phase boundary, separating the ordered ferromagnetic phase
from the high-temperature paramagnetic one. The phase-diagram line
separates the two phases of the model and intersects the
randomness axis at the critical value of the disorder strength.
Such qualitative sketching has been commonly used in most papers
for the
RFIM~\cite{newman:93,machta:00,newman:96,itakura:01,fytas:08b,aharony:78}
and closed form quantitative expressions are also known from the
early mean-field calculations~\cite{aharony:78}. However, it is
generally true that the quantitative aspects of phase diagrams
produced by mean-field treatments are very poor approximations.

On the theoretical side, a scaling picture is
available~\cite{villain:84,bray:85b,fisher:86b}. The
paramagnetic-ferromagnetic phase transition is ruled by a fixed
point [in the Renormalization-Group (RG) sense] at temperature
$T=0$~\cite{nattermann:98}. The spatial dimension $D$ is replaced
by $D-\theta$, in hyperscaling relations ($\theta\approx D/2$).
Nevertheless, one expects only two independent
exponents~\cite{aharony:76,schwartz:85,gofman:93,nattermann:98},
as in standard phase transitions~\cite{amit:05}. Unfortunately,
establishing the scaling picture is far from trivial. Perturbation
theory predicts that, in $D=3$, the ferromagnetic phase disappears
upon applying the tiniest random field~\cite{young:77}. Even if
the statement holds at all orders in perturbation
theory~\cite{parisi:79}, the ferromagnetic phase is stable in
$D=3$~\cite{bricmont:87}. Non-perturbative phenomena are obviously
at play~\cite{parisi:94,parisi:02}. Indeed, it has been suggested
that the scaling picture breaks down because of spontaneous
supersymmetry breaking, implying that more than two critical
exponents are needed to describe the phase
transition~\cite{tissier:11}.

On the experimental side, a particularly well researched
realization of the RFIM is the diluted antiferromagnet in an
applied magnetic field~\cite{belanger:98}. Yet, there are
inconsistencies in the determination of critical exponents. In
neutron scattering, different parameterizations of the scattering
line-shape yield mutually incompatible estimates of the thermal
critical exponent, namely $\nu=0.87(7)$~\cite{slanic:99} and
$\nu=1.20(5)$~\cite{ye:04}. Moreover, the anomalous dimension
$\eta=0.16(6)$~\cite{slanic:99} violates hyperscaling bounds, at
least if one believes experimental claims of a divergent specific
heat~\cite{belanger:83,belanger:98b}. Clearly, a reliable
parametrization of the line-shape would be welcome. This program
has been carried out for simpler, better understood
problems~\cite{martin-mayor:02}. Unfortunately, it is a common
belief that we do not have such a strong command over the RFIM
universality class.

The model has been also investigated by means of numerical
simulations~\cite{rieger:93,nowak:98,newman:96}. However, typical
Monte Carlo schemes get trapped into local minima with escape time
exponential in $\xi^\theta$, where $\xi$ denotes the correlation
length. Although sophisticated improvements have
appeared~\cite{fernandez:11b,fytas:08,vink:10,ahrens:13,picco:15},
these simulations produced low-accuracy data because they were
limited to linear sizes of the order of $L_{\rm max}\leq 32$.
Larger sizes can be achieved at $T=0$, through the well-known
mapping of the ground state to the maximum-flow optimization
problem~\cite{ogielski:86,auriac:86,sourlas:99,hartmann:95,auriac:97,swift:97,bastea:98,hartmann:99,hartmann:01,duxbury:01,middleton:01,middleton:02,dukovski:03,wu:05}.
Yet, $T=0$ simulations lack many tools, standard at $T>0$. In
fact, the numerical data at $T=0$ and their finite-size scaling
analysis mostly resulted in strong violations of
universality~\cite{sourlas:99,auriac:97,swift:97,hartmann:99}.

The criteria for determining the order of the low temperature
phase transition and its dependence on the form of the field
distribution have been discussed throughout the
years~\cite{aharony:78,galam:83,saxena:84,houghton:85,mattis:85,kaufman:86,sebastianes:87,arruda:89}.
In fact, different results have been proposed for different field
distributions, like the existence of a tricritical point at the
strong disorder regime of the system, present only in the bimodal
distribution~\cite{aharony:78,houghton:85}. Currently, despite the
huge efforts recorded in the literature, a clear picture of the
model's critical behavior is still lacking. Although the view that
the phase transition of the RFIM is of second-order is well
established~\cite{middleton:02,vink:10,fernandez:11b,fytas:08},
the extremely small value of the exponent $\beta$ continues to
cast some doubts. Moreover, a rather strong debate exists with
regards to the role of disorder: the available simulations are not
able to settle the question of whether the critical exponents
depend on the particular choice of the distribution for the random
fields, analogously to the mean-field theory
predictions~\cite{aharony:78}. Thus, the whole issue of the
model's critical behavior is under intense
investigation~\cite{tissier:11,fernandez:11b,fytas:08,ahrens:13,picco:15,hernandez:08,crokidakis:08,hadjiagapiou:11,akinci:11,picco:14}.

Recently, progress has been made towards this direction by the
present authors~\cite{fytas:13}. In particular, using a combined
approach of state of the art techniques from the pool of
statistical physics and graph theory, it was shown that the
universality class of the RFIM is independent of the form of the
implemented random-field distribution. This, somehow unexpected,
according to the current literature, result, was reached only
after a proper taming of the large scaling corrections, a fact
that, although emphasized many years ago~\cite{ogielski:86}, was
overlooked in numerous subsequent relevant investigations of the
model. In the current paper we present the full technical details
of our numerical implementation, originally outlined in
Ref.~\cite{fytas:13} and we provide some further numerical results
relevant to the scaling behavior of the specific heat and the
self-averaging aspects of the model in terms of the magnetic
susceptibility and the bond energy. We also discuss the scaling
aspects of the implemented maximum-flow algorithm.

The methods that we shall explain in the present paper will be
useful way beyond the context of the 3D RFIM. The most obvious
generalization is of course the RFIM in higher dimensions (see
e.g.~\cite{fytas:15}). However, similar ideas can be applied to
many disordered systems and should be useful when one needs to
take derivatives, or to perform reweighting extrapolations, with
respect to the disorder-distribution parameters. The ability to
obtain these derivatives is most important when the relevant RG
fixed-point lies at zero temperature (thus, parameters other than
temperature should be varied to cross the phase boundaries). For
instance, for 2D Ising spin glasses several RG fixed-points appear
at $T=0$ depending on the nature of the couplings
distribution~\cite{amoruso:03}. It should be possible then to
study the corresponding phase boundaries and RG flows using our
formalism. Another difficult problem that can be tackled with the
current prescription is the diluted antiferromagnet in a uniform
external field~\cite{belanger:98}. The ground state of this model
is degenerate, and it is thus difficult to sample with uniform
probability from the set of all ground states~\cite{bastea:98}.
Even if in experiments the external field is uniform, in
simulations it is desirable to add a small, local random noise to
the magnetic field~\cite{picco:15}. The small random magnetic
fields make it possible to employ the full formalism that we
derive in the following Sections. Furthermore, the
fluctuation-dissipation formulae elucidated below is also valid
when working at finite (rather than zero) temperature, which is
necessary for some algorithms~\cite{fernandez:11b, ahrens:13}.

The outline of paper is as follows: In the following
Sec.~\ref{section:model} we define the model and the random-field
distributions under study. In Sec.~\ref{section:zero_T_method} we
outline the $T=0$ maximum-flow algorithm, and in
Sec.~\ref{section:observables} we define the set of useful
physical observables that will be mainly analyzed. However a
complication arises: the sought observables cannot be
straightfowardly computed, as we explain in
Sec.~\ref{section:OLD-APPROACH}. The problems are overcome in
Sec.~\ref{section:FD}, where we derive explicitly a
fluctuation-dissipation formalism that allowed us to compute
connected and disconnected correlation functions from the $T=0$
data for each field distribution distinctively. The use of a
reweighting method with respect to the disorder strength consists
another asset at hand of our combinatorial scheme. In
Sec.~\ref{section:quotients} we give a brief description of our
finite-size scaling vehicle, the quotients
method~\cite{ballesteros:96}. In Sec.~\ref{section:results} and on
the basis of our main physical result of a single universality
class~\cite{fytas:13}, we illustrate the size evolution of several
effective critical exponents and we present a finite-size scaling
analysis of additional numerical data for the bond energy. For
this latter task, we adopt an extension of the quotients method,
necessary for monitoring the scaling of the effective exponent
$\alpha$ of the specific heat. Furthermore, we discuss the
self-averaging aspects of the model, by implementing a proper
noise to signal ratio for the magnetic susceptibility and the bond
energy, and we estimate the critical slowing-down exponent $z$ of
the zero-temperature algorithm used to generate the ground states
of the model. Our contribution ends with a summary in
Sec.~\ref{section:summary}.

\section{Model and random-field distributions}
\label{section:model}

Our $S_{x}=\pm 1$ spins are placed on a cubic lattice with size
$L$ and periodic boundary conditions. The Hamiltonian of the RFIM
in a general form may be written as
\begin{equation} \label{eq:H}
\mathcal{H}=-J\sum_{\langle
x,y\rangle}S_{x}S_{y}-\sum_{x}h_{x}S_{x},
\end{equation}
where in the above equation $J$ is the nearest-neighbors'
ferromagnetic interaction, which is set to be $J=1$. With $h_{x}$
we denote the set of independent quenched random fields. Common
field distributions considered in the literature are the Gaussian
and bimodal
distributions~\cite{belanger:91,rieger:95,nattermann:98}, for which
marginally distinct results have been
proposed~\cite{sourlas:99,auriac:97,swift:97,hartmann:99}.

In the current work  the quenched random fields $h_{x}$ are
extracted from one of the following double Gaussian (dG) or
Poissonian (P) distributions (with parameters $h_R$, $\sigma$):
\begin{eqnarray}
{\rm
  dG}^{(\sigma)}(h_{x};h_{R})&=&\frac{1}{\sqrt{8\pi\sigma^2}}\big[e^{-\frac{(h_{x}-h_{R})^{2}}{2\sigma^{2}}}+
  e^{-\frac{(h_{x}+h_{R})^{2}}{2\sigma^{2}}}\big],\ \ \label{eq:dGaussian}\\
{\rm P}(h_{x};\sigma)&=&\frac{1}{2|\sigma|} e^{-|h_{x}| /
\sigma}\,.\label{eq:Poisson}
\end{eqnarray}
The limiting cases $\sigma=0$ and $h_{R}=0$ of
Eq.~(\ref{eq:dGaussian}) correspond to the well-known bimodal (b)
and Gaussian (G) distributions, respectively. In the Poissonian
and Gaussian cases the strength of the random fields is
parameterized by $\sigma$, while in the double Gaussian case we
shall take $\sigma=1$ and $2$, and vary $h_R$.

As we are only interested in a $T=0$ study of the model by
estimating ground states via the use of efficient optimization
methods that will be discussed below, a proper choice of the
random-field distributions is of major importance in our task. In
particular, the main advantage of considering the double Gaussian
distribution of Eq.~(\ref{eq:dGaussian}) is that one can mimic for
certain values of $\sigma$ the double-peak structure of the
bimodal distribution, capturing its effects and at the same time
escaping the implication of non-degenerate ground states. As it is
well known, for cases of discrete distributions, like the bimodal,
degeneracy complicates the numerical solution of the system at
$T=0$, since one has to sweep over all the possible ground states
of the system~\cite{hartmann:95,bastea:98}. On the other hand, for
the cases of the above distributions~(\ref{eq:dGaussian}) and
(\ref{eq:Poisson}), the ground state of the system is
non-degenerate, so it is sufficient to calculate just one ground
state in order to get the necessary information.

\section{Zero-temperature algorithm}
\label{section:zero_T_method}

As already discussed extensively in the literature (see
Refs.~\cite{hartmann:04,hartmann:05} and references therein), the
RFIM captures essential features of models in statistical physics
that are controlled by disorder and have frustration. Such systems
show complex energy landscapes due to the presence of large
barriers that separate several meta-stable states. If such models
are studied using simulations mimicking the local dynamics of
physical processes, it takes an extremely long time to encounter
the exact ground state. However, there are cases where efficient
methods for finding the ground state can be utilized and,
fortunately, the RFIM is one such clear case. These methods escape
from the typical direct physical representation of the system, in
a way that extra degrees of freedom are introduced and an expanded
problem is finally solved. By expanding the configuration space
and choosing proper dynamics, the algorithm practically avoids the
need of overcoming large barriers that exist in the original
physical configuration space. An attractor state in the extended
space is found in time polynomial in the size of the system and
when the algorithm terminates, the relevant auxiliary fields can
be projected onto a physical configuration, which is the
guaranteed ground state.

The random field is a relevant perturbation at the pure fixed
point, and the random-field fixed point is at
$T=0$~\cite{villain:84,bray:85,fisher:86}. Hence, the critical
behavior is the same everywhere along the phase boundary and we
can predict it simply by staying at $T=0$ and crossing the phase
boundary at the critical field point. This is a convenient
approach because we can determine the ground states of the system
exactly using efficient optimization
algorithms~\cite{ogielski:86,auriac:86,sourlas:99,hartmann:95,auriac:97,swift:97,bastea:98,hartmann:99,hartmann:01,duxbury:01,middleton:01,middleton:02,dukovski:03,wu:05,fytas:13,alava:01,seppala:01,zumsande:08,shrivastav:11,ahrens:11,stevenson:11}
through an existing mapping of the ground state to the
maximum-flow optimization
problem~\cite{auriac:85,cormen:90,papadimitriou:94}. A clear
advantage of this approach is the ability to simulate large system
sizes and disorder ensembles in rather moderate computational
times. We should underline here that, even the most efficient
$T>0$ Monte Carlo schemes exhibit extremely slow dynamics in the
low-temperature phase of these
systems~\cite{hartmann:04,hartmann:05}. Further assets in the
$T=0$ approach are the absence of statistical errors and
equilibration problems, which, on the contrary, are the two major
drawbacks encountered in the $T>0$ simulation of systems with
rough free-energy landscapes~\cite{hartmann:04,hartmann:05}.

The application of maximum-flow algorithms to the RFIM is nowadays
well established~\cite{alava:01}. The most efficient network flow
algorithm used to solve the RFIM is the push-relabel algorithm of
Tarjan and Goldberg~\cite{goldberg:88}. For the interested reader,
general proofs and theorems on the push-relabel algorithm can be
found in standard textbooks~\cite{cormen:90,papadimitriou:94}. In
the present study we prepared our own C version of the algorithm
that involves a modification proposed by Middleton \emph{et
al.}~\cite{middleton:01,middleton:02,middleton:02b} that removes
the source and sink nodes, reducing memory usage and also
clarifying the physical
connection~\cite{middleton:02,middleton:02b}. For the sake of
completeness, we recall here the algorithm we use, which is
\emph{exactly} the algorithm proposed in
Refs.~\cite{middleton:01,middleton:02,middleton:02b}.

The algorithm starts by assigning an excess $x_i$ to each lattice
site $i$, with $x_i = h_i$. Residual capacity variables $r_{ij}$
between neighboring sites are initially set to $J$. A height
variable $u_i$ is then assigned to each node via a global update
step. In this global update, the value of $u_i$ at each site in
the set ${\cal T} =\left\{j|x_j<0\right\}$ of negative excess
sites is set to zero. Sites with $x_i \ge 0$ have $u_i$ set to the
length of the shortest path, via edges with positive capacity,
from $i$ to ${\cal T}$. The ground state is found by successively
rearranging the excesses $x_i$, via push operations, and updating
the heights, via relabel operations. When no more pushes or
relabels are possible, a final global update determines the ground
state, so that sites which are path connected by bonds with
$r_{ij}>0$ to ${\cal T}$ have $\sigma_i=-1$, while those which are
disconnected from ${\cal T}$ have $\sigma_i = 1$. A push operation
moves excess from a site $i$ to a lower height neighbor $j$, if
possible, that is, whenever $x_i>0$, $r_{ij} > 0$, and $u_j =
u_i-1$. In a push, the working variables are modified according to
$x_i \rightarrow x_i - \delta$, $x_j \rightarrow x_j + \delta$,
$r_{ij} \rightarrow r_{ij} - \delta$, and $r_{ji} \rightarrow
r_{ji} + \delta$, with $\delta = \min(x_i, r_{ij})$. Push
operations tend to move the positive excess towards sites in
${\cal T}$. When $x_i > 0$ and no further push is possible, the
site is relabelled, with $u_i$ increased to $1 + \min_{\{j| r_{ij}
> 0\}} u_j$. This is defined as a single push-relabel step;
the number of such steps will be our measure of algorithmic time.
In addition, if a set of highest sites ${\cal U}$ becomes
isolated, with $u_i > u_j+1$, for all $i\in{\cal U}$ and all
$j\notin{\cal U}$, the height $u_i$ for all $i\in{\cal U}$ is
increased to its maximum value, $L^{3}$, as these sites will
always be isolated from the negative excess nodes. The order in
which sites are considered is given by a queue. In this paper, we
have used the first-in-first-out (FIFO)
queue~\cite{middleton:02b}. The FIFO structure executes a
push-relabel step for the site $i$ at the front of a list. If any
neighboring site is made active by the push-relabel step, it is
added to the end of the list. If $i$ is still active after the
push-relabel step, it is also added to the end of the list. This
structure maintains and cycles through the set of active sites.
Last but not least, the computational efficiency of the algorithm
has been increased via the use of periodic global updates every
$L^{3}$ relabels~\cite{middleton:02,middleton:02b}.

\begin{table}
\caption{\label{tab:details} Summary of simulation details.}
\begin{ruledtabular}
\begin{tabular}{lccc}
Distribution  & $L_{\rm min}$ & $L_{\rm max}$ & $\mathcal{N}_{\rm samples}$ $(\times 10^{6})$\\
\hline
G                  &  $8$     &   192     &   10   \\
dG$^{(\sigma=1)}$  &  $8$     &   128     &   50    \\
dG$^{(\sigma=2)}$  &  $8$     &   128     &   10     \\
P                  &  $8$     &   192     &   10      \\
\end{tabular}
\end{ruledtabular}
\end{table}

Using the above version of the push-relabel algorithm, we
performed large-scale simulations of the RFIM defined above in
Eqs.~(\ref{eq:H}) - (\ref{eq:Poisson}) for a wide range of the
simulation parameters. Our tactic included three steps:
Originally, we performed preliminary runs with $\mathcal{N}_{\rm
samples}=10^{6}$, where $\mathcal{N}_{\rm samples}$ counts the
number of independent disorder realizations, to locate the
$h_{R}$- or $\sigma$-values (depending on the parametrization) of
the crossing points of the connected correlation length of the
system for pairs of lattice sizes of the form $(L,2L)$, as this is
indicated in the main heart of the scaling method used (see
below). Subsequently, the main part of the simulations took place
in these estimated crossing points, with details, in terms of
linear system sizes and disorder-averaged ensembles, summarized in
Table~\ref{tab:details}. In Table~\ref{tab:details} $L_{\rm
min}$($L_{\rm max}$) denotes the minimum(maximum) linear size
considered within the sequence of size points $L\in
\{8,12,16,24,32,48,64,96,128,192\}$. Finally, we performed an
additional set of simulations for triplets of systems sizes as
shown in Table~\ref{tab:specheat} in order to compute the critical
exponent of the specific heat via the scaling of the bond energy.
This will be exemplified in Sec.\ref{section:results}.

\section{Observables}
\label{section:observables}

An instance of the random fields $\{h_x\}$ is named a sample.
Thermal mean values are denoted as $\langle\cdots \rangle$, while
the subsequent average over samples is indicated by an over-line.
Two most basic quantities are the bond energy and the
order-parameter density:
\begin{equation}\label{eq:E_exch-m}
E_{\rm J}=-J\sum_{\langle x,y\rangle}S_{x}S_{y}\,,\quad
m=\frac{1}{L^D}\sum_x S_x\,.
\end{equation}

A crucial feature of the RFIM is that we have to deal with
\emph{two} different correlation functions, namely the
\emph{disconnected} and the {connected} propagators.

The disconnected propagator, is straightforward to compute both in
real, $G^{\rm (dis)}_{xy}$, and Fourier space, $\chi^{\rm
(dis)}_k$:
\begin{equation}\label{eq:G_diss}
G_{xy}^{\rm (dis)}=\overline{\langle S_{x}S_{y}\rangle}\,,\quad
\chi_k^{\rm (dis)}= L^{D} \overline{\langle |m_k|^2 \rangle}_k\,,
\end{equation}
where
\begin{equation}\label{eq:m_k}
m_k=\frac{1}{L^D} \sum_x \mathrm{e}^{\mathrm{i} k\cdot x} S_x\,.
\end{equation}
In particular, special notations are standard for vanishing
wavevector: $m_{k=(0,0,0)}=m$ (i.e. the order-parameter density),
and $\chi^{\rm (dis)}_{k=(0,0,0)}=\chi^{\rm (dis)}$ (i.e. the
disconnected susceptibility).

On the other hand, we have the connected propagator:
\begin{equation}
G_{xy}=\overline{\frac{\partial \langle S_{x}\rangle}{\partial
h_{y}}}\,.
\end{equation}
At finite temperature, one could compute $G_{xy}$ from the
Fluctuation-Dissipation Theorem
\begin{equation}\label{eq:FDT}
G_{xy}=\frac{1}{T}\overline{\langle S_xS_y\rangle- \langle
S_x\rangle\langle S_y\rangle}\,.
\end{equation}
However, we work directly at $T=0$, as explained in
Sec.~\ref{section:zero_T_method}. Therefore, Eq.~\eqref{eq:FDT} is
clearly unsuitable for us, and the methods of
Sec.~\ref{section:FD} will be needed (see also
Ref.~\cite{schwartz:85}). For later use, we note the symmetry
\begin{equation}
G_{xy}=G_{yx}=\frac{G_{xy}+G_{yx}}{2}\,.
\end{equation}
In fact, our numerical data will \emph{never} verify this symmetry
(because of statistical fluctuations), hence we prefer to use the
symmetrized propagator $(G_{xy}+G_{yx})/2$. Now, the connected
propagator in Fourier space is
\begin{equation}\label{eq:G-k}
\chi_k=\frac{1}{L^D}\sum_{x,y}\, \mathrm{e}^{\mathrm{i} k\cdot
(x-y)}\, \frac{G_{xy}+G_{yx}}{2}\,.
\end{equation}
Again, the case of vanishing wavevector deserves a special naming:
$\chi_{k=(0,0,0)}=\chi$ is the connected susceptibility.

From both propagators, we compute the connected, $\xi$, and
disconnected, $\xi^{\rm (dis)}$, second-moment correlation
lengths~\cite{amit:05,cooper:82}. Let
$k_\mathrm{min}=(2\pi/L,0,0)$, then
\begin{equation}
\xi^\#=\frac{1}{2\, \text{sin}
(\pi/L)}\sqrt{\frac{\chi^\#}{\chi^\#_{k_\mathrm{min}}}-1}\,,
\end{equation}
where the superscript $^\#$ stands both for the connected or the
disconnected case~\footnote{Due to a programming error, the
quantity denoted as $\xi$ in Ref.~\cite{fytas:13} did not coincide
with the standard definition of the second-moment correlation
length (see also Erratum of Ref.~\cite{fytas:13}). We would like
to point out that this error is corrected here and the results
shown in Figs.~\ref{fig:crossing} and \ref{fig:error_samples}
fully conform to the standard definition.}. Of course, we improve
our statistics by computing
$\chi^\#_{k_\mathrm{min}}=\frac{1}{3}\Big[\chi^\#_{k=(2\pi/L,0,0)}
+ \chi^\#_{k=(0,2\pi/L,0)}+\chi^\#_{k=(0,0,2\pi/L)}\Big]$.

Other important quantities are the well-known universal Binder
ratio
\begin{equation}\label{eq:U4-def}
U_4=\frac{\overline{\langle m^4\rangle}}{\overline{\langle
  m^2\rangle}^2}\,,
\end{equation}
and the susceptibilities ratio
\begin{equation}\label{eq:U22-def}
U_{22}=\frac{\chi^\mathrm{(dis)}}{\chi^{2}}\,,
\end{equation}
that we use as a platform for investigating the validity of the
so-called two-exponent scaling scenario, see
Sec.~\ref{section:results}.

\section{Problems with the straightforward approach}
\label{section:OLD-APPROACH}

Computing response functions is very important. Unfortunately, the
traditional approach for disordered systems (see
e.g.~\cite{ballesteros:98}) is not feasible at zero temperature.  The
problem is easily understood by considering the example of the Monte
Carlo computation of the magnetic susceptibility.

The traditional approach would start by generating
$\mathcal{N}_{\rm samples}$ of the random fields according
to the appropriate probability density $w(\{h_x\})$. Then, one
would add to each random field a uniform external field
\begin{equation}\label{eq:H-displaced}
h_x\longrightarrow h_x+H\,,
\end{equation}
and  the magnetic susceptibility would be estimated as
\begin{equation}\label{eq:bad-estimator}
\chi^{\mathrm{naive}} =\frac{1}{\mathcal{N}_{\rm samples}}
\sum_{s=1}^{\mathcal{N}_{\rm samples}} \left.\frac{\partial
\langle m_s\rangle_H}{\partial H}\right|_{H=0}\,,
\end{equation}
where $\langle m_s\rangle_H$ is the thermal expectation value of
instance $s$ under the displaced magnetic fields in
Eq.~\eqref{eq:H-displaced}. Yet, as we explain below, the naive
Monte Carlo estimator~\eqref{eq:bad-estimator} yields
$\chi^{\mathrm{naive}}=0$ with probability one for any smooth
random-field probability density $w(\{h_x\})$ such as ours, recall
Eqs.~(\ref{eq:dGaussian}) and (\ref{eq:Poisson}).

The approach outlined in Eq.~\eqref{eq:bad-estimator} fails
because, at zero temperature, the only spin-assignment with a
non-vanishing statistical weight is the ground state for the
Hamiltonian~\eqref{eq:H}.  The crucial point is that the ground
state is unique, excepting a zero-measure set in the
$L^D$-dimensional space spanned by the random-fields. Indeed,
consider two arbitrary but fixed spin-assignments, $\{S_x^{(1)}\}$
and $\{S_x^{(2)}\}$. The condition of equal energy
\begin{equation}\label{eq:hyper-plane}
\mathcal{H}(\{S_x^{(1)}\})=\mathcal{H}(\{S_x^{(2)}\})\,,
\end{equation}
defines an hyper-plane in the random-fields space. There are
$2^{L^D} (2^{L^D}-1)/2$ such space-dividing hyperplanes.  For
random fields $\{h_x\}$ not in these these hyper-planes, each of
the $2^{L^D}$ possible spin assignments has a distinct energy, and
thus the ground state is unique. Furthermore, the ordering of the
$2^{L^D}$ energy levels is fixed away from the hyper-planes (which
are the locus in random-fields space where level-crossings
happen).

Now, let us suppose that none of the $\mathcal{N}_{\rm samples}$
instances in Eq.~\eqref{eq:bad-estimator} lies \emph{exactly} in
one of the dividing hyper-planes [this happens with probability
one for any smooth $w(\{h_x\})$]. Then, for $H$ small enough, the
fields displacement in Eq.~(\ref{eq:H-displaced}) will not cross
any of the hyper-planes and thus, adding the field $H$ will leave
the ground state unvaried. In other words, $\left.\mathrm{d}
\langle m_s\rangle_H/\mathrm{d} H\right|_{H=0} =0$, with
probability one.

However, the connected susceptibility is~\emph{not} zero. The way
out of the paradox is simple: the $H$-derivatives in
Eq.~\eqref{eq:bad-estimator} are actually a sum of Dirac
$\delta$-functions, centered at the precise $H$ values that cause
the displaced fields \eqref{eq:H-displaced} to cross some of the
dividing hyper-planes~\eqref{eq:hyper-plane}. It is the integral
over the random-fields of these Dirac $\delta$-functions which
produces a finite susceptibility $\chi>0$:
\begin{equation}\label{eq:paradox}
\chi=\int\prod_x \mathrm{d} h_x\, w(\{h_x\})\ \left.\frac{\partial \langle m\rangle}{\partial H}\right|_{H=0}\,.
\end{equation}
We see the heart of the problem: naive Monte Carlo estimations
such as Eq.~\eqref{eq:bad-estimator} cannot correctly reproduce
integrals such as Eq.~\eqref{eq:paradox} when the integrand is a
such a singular object as a sum of Dirac's $\delta$-functions.

Nevertheless, people have tried to overcome the zero-measure
problem. For instance, one could keep $H$ finite and compute the Monte Carlo (MC) average
\begin{equation}
[ \langle m\rangle]^{\text{(MC)}}_H = \frac{1}{\mathcal{N}_{\rm
samples}} \sum_{s=1}^{\mathcal{N}_{\rm samples}} \langle
m_s\rangle_H\,,
\end{equation}
and then try to extrapolate to $H\to 0$ the slope $\mathrm{d} [
  \langle m\rangle]^{\text{(MC)}}_H/\mathrm{d} H$. Of course, the
smaller is $H$ the larger is the number $\mathcal{N}_{\rm samples}$ needed to observe some $H$-dependency.
Yet, reasonable tradeoffs between number of instances and size of
the applied field could be empirically found~\cite{hartmann:01}.

In Sec.~\ref{section:FD} we explain a completely different
approach that (i) allows to work directly at $H=0$ and (ii) avoids
Dirac's $\delta$-functions. How this is possible can be easily
understood by considering the following one-dimensional toy model.

\subsubsection{Toy model}

Imagine we have a single random field $h$. In analogy with the general
case, let us also assume that the magnetization, regarded as a
function of $h$, is constant but for a set of $R$ discontinuities:
\begin{equation}\label{eq:toy-1}
  \langle m \rangle_h = -1 +\sum_{i=1}^R\, [{m_{i+1} - m_i}]\, \theta(h-h_i)\,.
\end{equation}
In the above expression $\theta(x)$ is Heaviside step function,
$\theta(x>0)=1$ and $\theta(x<0)=0$, and the magnetization plateaux
are monotonically increasing, $m_{i+1} > m_i$, with $m_1=-1$
and $m_{R+1}=1$.

Now, if we displace the field, $h\longrightarrow h+H$, and take the
$H$ derivative in Eq.~\eqref{eq:toy-1}, a sum of Dirac
$\delta$-functions will arise, making unfeasible the Monte Carlo method.

However, it is useful to take one step back and recall how the
susceptibility is defined.  First, we consider the average
magnetization as a function of the displaced field
\begin{equation}\label{eq:toy-2}
\overline{\langle m \rangle}(H) =\int_{-\infty}^{\infty} \mathrm{d} w(h)\ \langle m \rangle_{h+H}\,.
\end{equation}
The derivative with respect to $H$ is taken only\emph{after} computing
the integral (the random-field probability density $w(h)$ must
decrease fast enough at infinity to make the integral
convergent). Yet, a change of variable $h'=h+H$ yields
\begin{equation}\label{eq:toy-3}
\overline{\langle m \rangle}(H) =\int_{-\infty}^{\infty} \mathrm{d} w(h-H)\ \langle m \rangle_{h}\,.
\end{equation}
The change of variable is mathematically sound, as it relies only on
the translational invariance of the integration measure in
Eq.~\eqref{eq:toy-2}. If the probability density $w(h)$ is smooth, one can now interchange derivative and integral obtaining
\begin{equation}\label{eq:toy-4}
\chi_{\text{toy model}}= \int_{-\infty}^{\infty} \mathrm{d} w(h)\
\frac{-1}{w(h)}\frac{\mathrm{d} w}{\mathrm{d} h}\,  \langle m
\rangle_{h}\,.
\end{equation}
The integrand in Eq.~\eqref{eq:toy-4} is a regular function,
allowing for a Monte Carlo estimation of the form
\begin{equation}\label{eq:toy-5}
\chi_{\text{toy model}}^{\rm (MC)}=\frac{1}{\mathcal{N}_{\rm
samples}} \sum_{s=1}^{\mathcal{N}_{\rm samples}}\ \langle m
\rangle_{h_s}\left.\frac{-1}{w(h_s)}\frac{\mathrm{d} w}{\mathrm{d}
h}\right|_{h=h_s}\,,
\end{equation}
where the independent random fields $h_s$ are obtained with weight
$w(h)$.  Note that the summands in Eq.~\eqref{eq:toy-5}
\emph{cannot} be interpreted as the magnetic susceptibility of a
given instance (there are no Dirac $\delta$-functions). However,
$\chi_{\text{toy model}}^{\rm (MC)}$ does converge to
$\chi_{\text{toy model}}$ in the limit of large $\mathcal{N}_{\rm
samples}$.

\section{Fluctuation-dissipation formalism}
\label{section:FD}

Reweighting methods are a major asset for numerical studies of
critical phenomena~\cite{falcioni:82,ferrenberg:88}: From a
\emph{single} simulation at a given temperature we get a
\emph{continuous} curve for (say) the disconnected susceptibility,
$\chi^{\mathrm{(dis)}}(T)$.

However, we will be working at zero temperature. Hence, standard
reweighting methods are not useful for us. In fact, we shall
explain here our extension of reweighting methods originally
devised for percolation
studies~\cite{harris:94,ballesteros:98,ballesteros:98b}. From a
single simulation, we extrapolate the mean value of observables to
nearby parameters of the disorder distribution. We varied $\sigma$
for the Poissonian and Gaussian distributions, see panel (a) in
Fig.~\ref{fig:crossing} below for an illustrative flavor, and
$h_{R}$ for the double Gaussian distribution. These reweighting
methods were instrumental for our previous work~\cite{fytas:13}.

As we discuss below, a closely related problem is the computation of
the connected correlation functions (recall also
Sect.~\ref{section:OLD-APPROACH}). Our solution for the case of the
Gaussian distribution, in Sec.~\ref{subsect:FD-Gaussian}, will turn
out to be identical to the one in Ref.~\cite{schwartz:85}. However,
modifications are needed for the Poissonian or double Gaussian
distributions, which are explained in
Secs.~\ref{subsect:FD-Poissonian} and~\ref{subsect:FD-dGaussian},
respectively.

For all three distributions, we shall compute the connected
propagator by adding a \emph{source} $\tilde h_x$ to the random
fields:
\begin{equation}\label{eq:h_sources}
h_x\longrightarrow h_x+\epsilon \tilde h_x,
\end{equation}
where $\epsilon$ is a small parameter.  At variance with the
random fields $\{ h_x\}$, the sources $\{\tilde h_x\}$ will be
arbitrary but \emph{fixed}: the over-line will indicate and
average only with respect to the $\{ h_x\}$. Then, the connected
propagator $G_{xy}(=G_{yx})$ follows from the Taylor expansion
\begin{equation}\label{eq:FD-INTEGRAL}
\overline{\langle S_y \rangle_{\{ h_x+\epsilon \tilde h_x\}}}=
\overline{\langle S_y \rangle} + \epsilon \sum_x G_{yx} \tilde h_x
+{\cal O}(\epsilon^2)\,.
\end{equation}
In the above expression, $\langle S_y \rangle_{\{ h_x+\epsilon
\tilde h_x\}}$ is the thermal expectation value obtained when
plugging $\{ h_x+\epsilon \tilde h_x\}$ as the random magnetic
fields in the Hamiltonian, Eq.~\eqref{eq:H}.

The formalism will be explained in the same way, for all three
random-field distributions.  We start from the general observation
that computing thermal expectation values at $T=0$ is trivial: one
just needs to evaluate the function of interest, see
Sec.~\ref{section:observables}, on the ground-state spin
assignment corresponding to a given sample (recall that a sample
is characterized by a set of random fields $\{h_x\}$). In this
sense, thermal mean-values are mere functions of the $\{h_x\}$.
Next, we observe that a special function of the random-fields,
when averaged over the $\{h_x\}$, is equal to the connected
propagator. Finally, we show how to perform reweighting
extrapolations for a generic function of the random fields ${\cal
F}(\{h_x\})$.

Before we start, let us mention that a practical consideration had
an important impact in the designing of our
Fluctuation-Dissipation formalism. We simulated a large number of
samples ($\sim 10^7$) on large system sizes ($L=192$), see
Table~\ref{tab:details}. Clearly, storing in the hard drive all
the corresponding ground-state assignments is out of the question.
Therefore, we need to select \emph{beforehand} a small set of
quantities to be computed on the ground-state spin assignment and
stored on the hard drive. This small set of observables includes
$E_{\rm J}$, $m$ and $m_{k_\mathrm{min}}$, recall
Sec.~\ref{section:observables}, but also the quantities needed to
compute the connected propagators and the reweighting
extrapolations (in all cases, we restricted the wavevectors to a
bare minimum: $k=(0,0,0)$ and $k=k_{\mathrm{min}}$).

\subsection{Gaussian distribution} \label{subsect:FD-Gaussian}

The combined probability density for our $N=L^D$  Gaussian random
fields with width parameter $\sigma$ is
\begin{equation}\label{eq:w-G}
w^{\mathrm{G}}(\{h_x\},\sigma)=\frac{1}{(2\pi\sigma^2)^{\frac{N}{2}}}\,
\mathrm{e}^{-\frac{1}{2\sigma^2}\sum_x h_x^2}\,.
\end{equation}
Our computation starts from Eq.~\eqref{eq:FD-INTEGRAL}:
\begin{eqnarray}
&&\overline{\langle S_y \rangle_{\{ h_x+\epsilon \tilde h_x\}}}=\nonumber\\
&&= \int \prod_x \mathrm{d}h_x\, w^{\mathrm{G}}(\{h_x\},\sigma)\,  \langle S_y \rangle_{\{ h_x+\epsilon \tilde h_x\}}\label{eq:connectG-1}\\
&&=\int \prod_x \mathrm{d}h'_x\, w^{\mathrm{G}}(\{h'_x-\epsilon
\tilde h_x\},\sigma)\,  \langle S_y \rangle_{\{
h'_x\}}\,.\label{eq:connectG-2}
\end{eqnarray}
In the above expressions the $N$ integrals extend from $-\infty$
to $+\infty$.  We went from~\eqref{eq:connectG-1}
to~\eqref{eq:connectG-2} by changing integration variables as
$h'_x= h_x+\epsilon \tilde h_x$ [we shall drop the prime for the
dummy
  integration variables, $h_x'$ in Eq.~\eqref{eq:connectG-2}]. Now,
one just needs to Taylor-expand in the small parameter $\epsilon$
in Eq.~\eqref{eq:connectG-2}. A direct comparison with
Eq.~\eqref{eq:FD-INTEGRAL} yields
\begin{eqnarray}
G_{zy}&=&
\int \prod_x \mathrm{d}h_x\, w^{\mathrm{G}}(\{h_x\},\sigma)\, \frac{h_z \langle S_y \rangle_{\{ h_x\}}}{\sigma^2}\\
&=& \frac{\overline{ h_z \langle
S_y\rangle}}{\sigma^2}\,.\label{eq:connectG-final}
\end{eqnarray}
We now use Eq.~\eqref{eq:G-k} to compute the propagator in the
Fourier space
\begin{equation}\label{eq:chi-G}
\chi_k = L^D \frac{\overline{ \langle h^{\mathrm{G}}_{-k} m_k +
h^{\mathrm{G}}_k m_{-k} \rangle }}{2 \sigma^2}\,,
\end{equation}
where $m_k$ was defined in Eq.~\eqref{eq:m_k} and
\begin{equation}\label{eq:h_k_G}
h^{\mathrm{G}}_k=\frac{1}{L^D} \sum_x \mathrm{e}^{\mathrm{i}
k\cdot x} h_x\,.
\end{equation}

The reader will note that Eq.~\eqref{eq:chi-G} was obtained in
Ref.~\cite{schwartz:85} (yet, our argument is sound as well when
one starts directly at $T=0$, which is exactly our case). Our
rationale for recalling this fluctuation-dissipation argument here
is that the derivation of the new formulae in
Secs.~\ref{subsect:FD-Poissonian} and~\ref{subsect:FD-dGaussian}
is completely analogous.

At this point it should be obvious that, for all the observables
of interest, we are after the computation of multidimensional
integrals of the form
\begin{equation}\label{eq:RWG-0}
\left.\overline{\cal F}\right|_{\sigma} = \int \prod_x
\mathrm{d}h_x\, w^{\mathrm{G}}(\{h_x\},\sigma)\, {\cal
F}(\{h_x\})\,,
\end{equation}
where ${\cal F}(\{h_x\})$ could be $ {\cal F}=\langle S_z S_y
\rangle_{\{
  h_x\}}$, or $ {\cal F}= h_z \langle S_y \rangle_{\{
  h_x\}}$, etc. Now, we need to solve three problems:
\begin{enumerate}
\item Compute derivatives with respect to $\sigma$, $D_\sigma
\overline{\cal F}$. Recall that $\sigma$ is the width for the
Gaussian weight in Eq.~\eqref{eq:RWG-0}. \item Extrapolate the
expectation values at $\sigma+\delta \sigma$ from integrals at
$\sigma$ such as Eq.~\eqref{eq:RWG-0}. \item Estimate how large
the extrapolation window $\delta \sigma$ may
  be in a numerical simulation.
\end{enumerate}
Fortunately, we can solve all three problems with a single trick.
The starting point is
\begin{eqnarray}
\left.\overline{\cal F}\right|_{\sigma+\delta\sigma} & = & \int
\prod_x \mathrm{d}h_x\,
w^{\mathrm{G}}(\{h_x\},\sigma+\delta\sigma)\,
 {\cal F}(\{h_x\})\,, \label{eq:RWG-1}\\
&=&
\int \prod_x \mathrm{d}h_x\, w^{\mathrm{G}}(\{h_x\},\sigma)\, \times \nonumber\\
&\times&{\cal F}(\{h_x\}) {\cal R}(\{h_x\},\sigma,\delta\sigma)\,,
\label{eq:RWG-2}
\end{eqnarray}
where the reweighting factor ${\cal R}$ is just the ratio of
probability densities:
\begin{eqnarray}
{\cal R}(\{h_x\},\sigma,\delta\sigma) = \frac{w^{\mathrm{G}}(\{h_x\},\sigma+\delta\sigma)}{w^{\mathrm{G}}(\{h_x\},\sigma)}&&\\
= \left(\frac{\sigma}{\sigma+\delta\sigma}\right)^N\,
\mathrm{e}^{\frac{1}{2}\Big[\sigma^{-2}-(\sigma+\delta\sigma)^{-2}\Big]\sum_x
h_x^2}\,.&&\label{eq:RWF-G}
\end{eqnarray}

The computation of $\sigma$-derivatives follows straightforwardly
by Taylor expanding the reweighting factor in $\delta \sigma$:
\begin{equation}
{\cal R}(\{h_x\},\sigma,\delta\sigma+\epsilon)={\cal
R}(\{h_x\},\sigma,\delta\sigma) \Big(1+ \epsilon {\cal D} + {\cal
O} (\epsilon^2)\Big)\,,
\end{equation}
where
\begin{equation}
{\cal D}
(\{h_x\},\sigma,\delta\sigma)=\frac{1}{\sigma+\delta\sigma}\Big[\frac{\sum_x
h_x^2}{(\sigma+\delta \sigma)^2} - N\Big]\,.
\end{equation}

Our reweighting formulae can be cast in a more aesthetically
appealing form
\begin{equation}\label{eq:RWG-3}
\left.\overline{{\cal F}}\right|_{\sigma+\delta\sigma}=
\left.\overline{{\cal F} {\cal
R}_{\sigma,\delta\sigma}}\right|_{\sigma}\,,\
\left.D_\sigma\overline{{\cal
F}}\right|_{\sigma+\delta\sigma}=\left.\overline{{\cal F} {\cal
R}_{\sigma,\delta\sigma} {\cal
D}_{\sigma,\delta\sigma}}\right|_{\sigma}\,.
\end{equation}
Note that Eq.~\eqref{eq:RWG-3} refers to a function ${\cal F}$ of
the random-fields \emph{only}. Explicit dependency on $\sigma$,
like in $G_{zy}=\overline{h_z\langle S_y
  \rangle}/\sigma^2$, is not included but can be taken care of
straightforwardly.

In summary, in order to perform a complete reweighting study for
each sample we need to store on the hard drive only $E_{\rm J}$,
$m_k$, $h^{\mathrm{G}}_{-k}m_{k}+h^{\mathrm{G}}_{k}m_{-k}$
(restricting ourselves to $k=(0,0,0)$ and $k=k_{\mathrm{min}}$),
as well as $\sum_x h_x^2$.

The final question we need to address is: how large $\delta
\sigma$ can reasonably be in a Monte Carlo simulation? Of course,
the question is ill-posed, because the answer depends on how many
samples are simulated. In the limit of infinite statistics, one
could have arbitrarily large $\delta \sigma$. However, this ideal
situation is never reached in practice.  As a rule of thumb one
may use many different criteria, but all of them boil down to
requiring that the \emph{typical} set of random-fields for
$\sigma+\delta \sigma$ could also be typical at $\sigma$ (or, at
least, not too unusual). A particularly simple such criterium
requires the absolute value of
\begin{equation}
\left.\overline{\sum_x h_x^2}\right|_{\sigma+\delta\sigma} -
\left.\overline{\sum_x h_x^2}\right|_{\sigma}= N
[(\sigma+\delta\sigma)^2-\sigma^2]\,
\end{equation}
to be no larger than the dispersion of $\sum_x h_x^2$ at $\sigma$,
namely $\sqrt{2 N}\sigma^2$. The resulting bound is
\begin{equation}
|\delta\sigma|\leq \sqrt{\frac{\sigma^2}{2 N}}\,.
\end{equation}

\subsection{Poissonian distribution}
\label{subsect:FD-Poissonian}

This case is a straightforward translation of the results in
Sec.~\ref{subsect:FD-Gaussian}. Since there is not any new idea
involved, let us just give the main results.

The connected propagator in real space is
\begin{equation}\label{eq:connectP-final}
G_{zy}=\frac{\overline{ h_z \langle S_y\rangle}}{|h_z| \sigma}\,.
\end{equation}
Note the small, but crucial, difference with
Eq.~\eqref{eq:connectG-final}: we correlate $\langle S_y\rangle $
with the \emph{sign} of $h_z$. In the Fourier space,
Eq.~\eqref{eq:connectG-final} translates to
\begin{equation}\label{eq:chi-P}
\chi_k = L^D \frac{\overline{ \langle h^{\mathrm{P}}_{-k} m_k +
h^{\mathrm{P}}_k m_{-k} \rangle }}{2 \sigma}\,,
\end{equation}
where $m_k$ was defined in Eq.~\eqref{eq:m_k} and
\begin{equation}\label{eq:h_k_P}
h^{\mathrm{P}}_k=\frac{1}{L^D} \sum_x \mathrm{e}^{\mathrm{i}
k\cdot x} \frac{h_x}{|h_x|}\,.
\end{equation}
Note, again, that we Fourier-transform the sign of the Poissonian
random fields.

The reweighting factor is again the ratio of probability densities
for the Poisson fields:
\begin{eqnarray}
&&{\cal R}(\{h_x\},\sigma,\delta\sigma) = \\
&&= \left(\frac{\sigma}{\sigma+\delta\sigma}\right)^N\,
\mathrm{e}^{\Big[\sigma^{-1}-(\sigma+\delta\sigma)^{-1}\Big]\sum_x
|h_x|}\,,
\end{eqnarray}
and the derivative operator follows from a Taylor expansion with
respect to $\delta\sigma$:
\begin{equation}
{\cal R}(\{h_x\},\sigma,\delta\sigma+\epsilon)={\cal
R}(\{h_x\},\sigma,\delta\sigma) \Big(1+ \epsilon {\cal D} + {\cal
O} (\epsilon^2)\Big)\,,
\end{equation}
where
\begin{equation}
{\cal D}
(\{h_x\},\sigma,\delta\sigma)=\frac{1}{\sigma+\delta\sigma}\Big[\frac{\sum_x
|h_x|}{(\sigma+\delta \sigma)} - N\Big]\,.
\end{equation}

The final reweighting formulae can be cast in exactly the same
form that we found for the Gaussian random fields
\begin{equation}\label{eq:RWG-P}
\left.\overline{{\cal F}}\right|_{\sigma+\delta\sigma}=
\left.\overline{{\cal F} {\cal
R}_{\sigma,\delta\sigma}}\right|_{\sigma}\,,\
\left.D_\sigma\overline{{\cal
F}}\right|_{\sigma+\delta\sigma}=\left.\overline{{\cal F} {\cal
R}_{\sigma,\delta\sigma} {\cal
D}_{\sigma,\delta\sigma}}\right|_{\sigma}\,.
\end{equation}

As for the maximum reasonable reweighting extrapolation, we also
use an analogous criterium: The absolute value of the difference
\begin{equation}
\left.\overline{\sum_x |h_x|}\right|_{\sigma+\delta\sigma} -
\left.\overline{\sum_x |h_x|}\right|_{\sigma}=
N(\sigma+\delta\sigma)-N\sigma\,
\end{equation}
should be no larger than the dispersion of $\sum_x |h_x|$ at
$\sigma$, namely $\sqrt{N}\sigma$. The resulting bound is
\begin{equation}
|\delta\sigma|\leq \sqrt{\frac{\sigma^2}{N}}\,.
\end{equation}

\subsection{Double Gaussian distribution}
\label{subsect:FD-dGaussian}

Our formalism for this distribution is slightly more complicated.
Let us start by explaining how we obtain a random field
distributed as prescribed in Eq.~\eqref{eq:dGaussian}. For each
$h_x$ we extract two independent random variables. One of them is
discrete, $\eta_x=\pm 1$, with $50\%$ probability. The other
variable, $\varphi_x$, is gaussian distributed with zero mean and
unit variance. Then, we set [the width $\sigma$ is given in
Eq.~\eqref{eq:dGaussian}]
\begin{equation}\label{eq:dG-two-variables}
h_x=h_R\eta_x+\sigma \varphi_x\,.
\end{equation}

The combined probability density for our $2N$ variables is
\begin{equation}\label{eq:dG}
w^{\mathrm{dG}}(\{\eta_x,\varphi_x\})=\frac{\mathrm{e}^{-\frac{1}{2}\sum_x
\varphi_x^2}}{2^N (2\pi)^{N/2}}\,.
\end{equation}

In order to understand the origin of the additional complications
for this distribution, let us add a source (i.e.
$h_x\longrightarrow h_x+\epsilon \tilde h_x$), while we
simultaneously modify the position of the peaks, (i.e.
$h_R\longrightarrow h_R+\delta h_R$)~\footnote{The
  reader may check that this procedure is inconsequential for the
  Gaussian or the Poissonian distributions. For instance, in the
  Gaussian case, the analogous of Eq.~\eqref{eq:connectG-final}
  obtained by modifying simultaneously the width of the distribution,
  $\sigma\longrightarrow \sigma+\delta\sigma$, would be
  $\left. G_{zy}\right|_{\sigma+\delta\sigma}=\left.\overline{h_z\langle
    S_y\rangle {\cal R}^{\mathrm
      {G}}_{\sigma,\delta\sigma}}\right|_{\sigma}/(\sigma+\delta\sigma)^2$, where
${\cal R}^\mathrm{G}_{\sigma,\delta\sigma}$ is the reweighting
factor appropriated for the Gaussian distribution, see
Eq.~\eqref{eq:RWF-G}. This is exactly the same result that one
obtains by combining Eqs.~\eqref{eq:connectG-final}
and~\eqref{eq:RWG-3}.}.

Under our circumstances, Eq.~\eqref{eq:FD-INTEGRAL} reads
\begin{eqnarray}
&&\left.\overline{\langle S_y \rangle_{\{ h_x+\epsilon \tilde h_x\}}}\right|_{h_R+\delta h_R}=\nonumber\\
&&= \sum_{\{\eta_x\}}\int \prod_x \mathrm{d}\varphi_x\, w^{\mathrm{dG}}(\{\eta_x,\varphi_x\})\,  \langle S_y \rangle_{\{\hat h_x\}}\,,\label{eq:connectdG-1}\\
&&\hat h_x=h_R\eta_x+\delta h_R \eta_x+\sigma \varphi_x+\epsilon
\tilde h_x\,.\label{eq:dG-aux-1}
\end{eqnarray}
The sum in Eq.~\eqref{eq:connectdG-1} extends to the $2^N$
possible values of the discrete variables $\eta_x$. The problem
now is apparent from Eq.~\eqref{eq:dG-aux-1}. For each site, we
have a single integration variable, namely $\varphi_x$. We need to
carry out a change of variable that brings Eq.~\eqref{eq:dG-aux-1}
to the form in Eq.~\eqref{eq:dG-two-variables}:
\begin{equation}\label{eq:dG-change-of.variable}
\varphi'_x=\varphi_x + \frac{\delta h_R \eta_x +\epsilon \tilde
h_x}{\sigma}\,.
\end{equation}
In other words, if $\delta h_R\neq 0$, there is no way of
distinguishing $\delta h_R\eta_x$ from the source term $\epsilon
\tilde h_x$.

With this caveat in mind, and dropping the prime in the dummy
integration variables, Eq.~\eqref{eq:connectdG-1} now reads
\begin{eqnarray}
&&\left.\overline{\langle S_y \rangle_{\{ h_x+\epsilon \tilde h_x\}}}\right|_{h_R+\delta h_R}=\nonumber\\
&&= \sum_{\{\eta_x\}}\int \prod_x \mathrm{d}\varphi_x\,
w^{\mathrm{dG}}\Big(\big\{\eta_x,\varphi_x-\frac{\delta h_R \eta_x
+\epsilon \tilde h_x}{\sigma}\big\}\Big)\,
\times\nonumber\\
&&\quad\quad\quad\times\, \langle S_y \rangle_{\{\hat h_x\}}\,,\label{eq:connectdG-2}\\
&&\hat h_x=h_R\eta_x+\sigma \varphi_x\,.\label{eq:dG-aux-2}
\end{eqnarray}
Now,
\begin{eqnarray}\label{eq:dG-aux-3}
&&w^{\mathrm{dG}}\Big(\big\{\eta_x,\varphi_x-\frac{\delta h_R \eta_x +\epsilon \tilde h_x}{\sigma}\big\}\Big)=\\[1mm]
&&= w^{\mathrm{dG}}(\{\eta_x,\varphi_x\})\, {\cal R}(\{\eta_x,\varphi_x\},\delta h_R)\,\times\nonumber\\[1mm]
&&\times\,\mathrm{exp}\bigg[\frac{\epsilon}{\sigma}\sum_x \tilde
h_x\bigg(\varphi_x-\frac{\delta h_R\eta_x}{\sigma}\bigg)\bigg]\,\times\nonumber\\[1mm]
&&\times\,\mathrm{exp}\bigg[-\frac{\epsilon^2}{2
\sigma^2}\sum_x \tilde h_x^2\bigg]\,,\nonumber
\end{eqnarray}
where the reweighting factor appropriate for our implementation of
the double Gaussian distribution is
\begin{eqnarray}
{\cal R}(\{\eta_x,\varphi_x\},\delta h_R)&=&\mathrm{exp}\bigg[\frac{\delta
h_R}{\sigma}\sum_x \eta_x\varphi_x\bigg]\,\times\nonumber\\[1mm]
&\times&\mathrm{exp}\bigg[-\frac{\delta h_R^2
N}{2\sigma^2}\bigg]\,.
\end{eqnarray}

Taylor-expanding with respect to $\epsilon$ in
Eq.~\eqref{eq:dG-aux-3} we finally get the connected propagator
\begin{equation}
\left.G_{zy}\right|_{h_R+\delta
  h_R}=\frac{1}{\sigma}\left.\overline{{\cal R}_{\delta h_R}
  \big(\varphi_z -\frac{\delta h_R\eta_z}{\sigma}\big) \langle S_y\rangle}\right|_{h_R}\,.
\end{equation}
In particular, the correction term $-\delta h_R \eta_z$ was absent
for the Poissonian and Gaussian distributions. Similarly, one can
get the connected propagator in the Fourier space
\begin{eqnarray}\label{eq:chi-dG}
\left.\chi_k\right|_{h_R+\delta h_R} = &\frac{L^D}{\sigma}& \overline{{\cal R}_{\delta h_R} \bigg\langle \bigg(\hat\varphi_{-k}-\frac{\delta h_R\hat\eta_-k}{\sigma}\bigg)\, m_k\, +}\nonumber\\
&+&\left.\overline{ \bigg(\hat\varphi_{k}-\frac{\delta h_R\hat\eta_k}{\sigma}\bigg)\,
m_{-k}\, \bigg\rangle}\right|_{h_R}\,,
\end{eqnarray}
where $m_k$ was defined in Eq.~\eqref{eq:m_k} and
\begin{equation}\label{eq:h_k_d_G}
\hat\varphi_k=\frac{1}{L^D} \sum_x \mathrm{e}^{\mathrm{i} k\cdot
x} \varphi_x\quad,\quad \hat\eta_k=\frac{1}{L^D} \sum_x
\mathrm{e}^{\mathrm{i} k\cdot x} \eta_x\,.
\end{equation}

\begin{figure}[htbp]
\includegraphics*[width=8 cm]{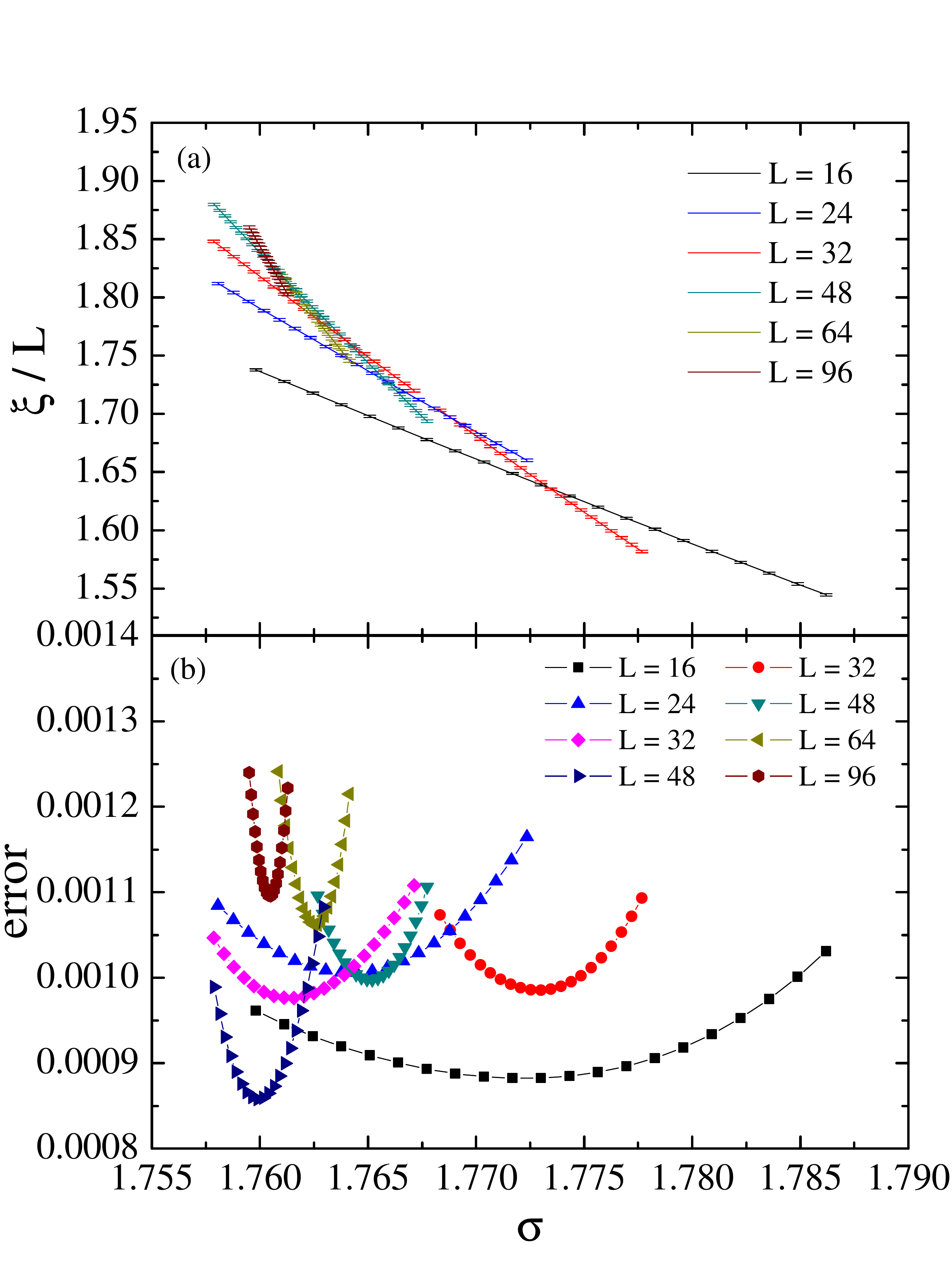}
\caption{\label{fig:crossing} (Color online) (a) For several
system sizes, we show  $\xi / L$ as a function of the strength of
the Poissonian random field $\sigma$. Lines join data obtained
from reweighting
  extrapolation (discontinuous lines of the same
  color come from independent simulations).  In the large-$L$ limit, $\xi/L$
  is $L$-independent at the critical point $\sigma^{\rm (c)}$. In the quotients
  method, we consider the $\xi/L$ curves for pair of lattices $(L,2L)$ and
  seek the $\sigma$ where they cross. This crossing is employed for
  computing effective, $L$-dependent critical exponents with
  Eq.~\eqref{eq:QO}. (b) Illustration of statistical errors in the universal
 ratio $\xi / L$ for the pairs of the system sizes shown in panel (a).}
\end{figure}

Instead for disconnected observables (e.g. $E_{\rm J}$, connected
propagators, etc.) the reweighting formulae take the standard form
\begin{equation}\label{eq:RWG-dG}
\left.\overline{{\cal F}}\right|_{\sigma+\delta\sigma}=
\left.\overline{{\cal F} {\cal R}_{\delta
h_R}}\right|_{\sigma}\,,\ \left.D_\sigma\overline{{\cal
F}}\right|_{\sigma+\delta\sigma}=\left.\overline{{\cal F} {\cal
R}_{\delta h_R} {\cal D}_{\delta h_R}}\right|_{\sigma}\,,
\end{equation}
where, in this case, the derivative operator is
\begin{equation}
{\cal D}(\{\eta_x,\varphi_x\},\delta h_R)=\frac{1}{\sigma}\sum_x
\big(\eta_x\phi_x -\frac{\delta h_R}{\sigma}\big)\,.
\end{equation}

Finally, we need to asses the maximum sensible size for $\delta
h_R$. The simplest way to proceed is to compute the moments of the
reweighting factor
\begin{equation}
\overline{{\cal R}_{\delta
h_R}^k}=\mathrm{exp}\bigg[N\frac{k^2-k}{2}\frac{\delta
h_R^2}{\sigma^2}\bigg]\,.
\end{equation}
\begin{figure}[htbp]
\includegraphics*[width=9 cm]{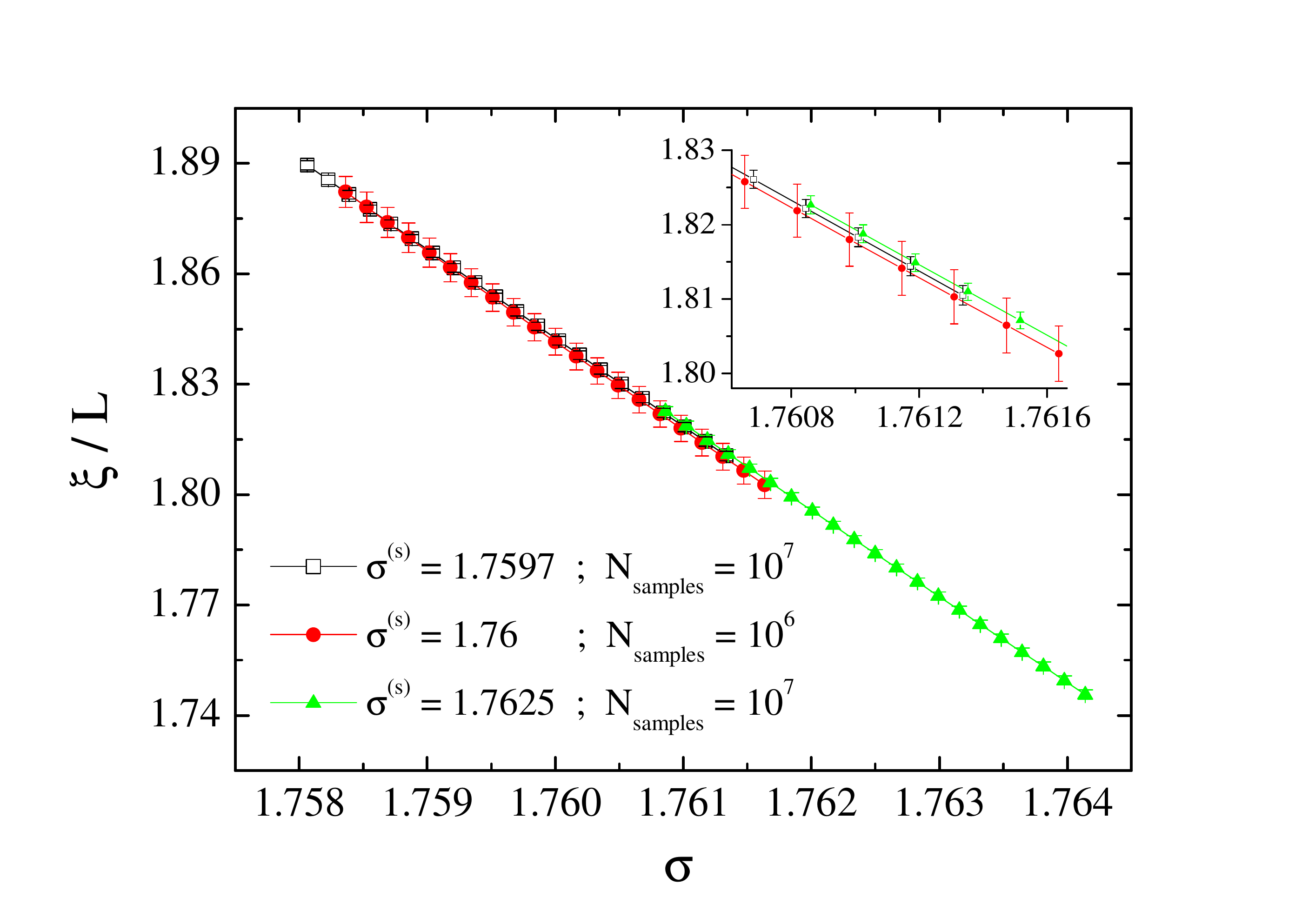}
\caption{\label{fig:error_samples} (Color online) Universal ratio
$\xi / L$ of an $L = 64$ Poissonian RFIM for three different sets
of simulations, corresponding to different simulation values
$\sigma^{\rm (s)}$ and different sets of random realizations. The
inset is a mere enlargement of the intermediate regime of $\sigma$
values.}
\end{figure}
If we now demand the dispersion (i.e., square-root of variance) to
be as large as the mean-value, we get
\begin{equation}
|\delta h_R|\leq \sigma\sqrt{\frac{\log 2}{N}}\,.
\end{equation}

\section{Quotients method}
\label{section:quotients}

To extract the values of critical points, critical exponents, and
dimensionless universal quantities, we employed the quotients
method, also known as phenomenological
renormalization~\cite{ballesteros:96,amit:05,nightingale:76}. This
method allows for a particularly transparent study of corrections
to scaling, that up to now have been considered as the Achilles'
heel in the study of the $D\geq 3$ random-field problem. We should
note that previous applications of the method include diluted
antiferromagnets~\cite{fernandez:11b} and the spin-glass problem,
see Ref.~\cite{janus:13} and references therein.

We compare observables computed in pair of lattices $(L,2L)$. We
start imposing scale-invariance by seeking the $L$-dependent
critical point: the value of $\sigma$ ($h_R$ for the dG), such
that $\xi_{2L}/\xi_L=2$ (i.e. the crossing point for $\xi_L/L$,
see Fig.~\ref{fig:crossing}(a)). Now, for dimensionful quantities
$O$, scaling in the thermodynamic limit as $\xi^{x_O/\nu}$, we
consider the quotient $Q_O=O_{2L}/O_L$ at the crossing. For
dimensionless magnitudes $g$, we focus on $g_{2L}$. In either
case, one has:
\begin{equation}\label{eq:QO}
Q_O^\mathrm{(cross)}=2^{x_O/\nu}+\mathcal{O}(L^{-\omega})\,,\
g_{(2L)}^\mathrm{(cross)}=g^\ast+\mathcal{O}(L^{-\omega})\,,
\end{equation}
where $x_O/\nu$, $g^\ast$ and the scaling-corrections exponent
$\omega$ are universal.

\begin{figure}[htbp]
\includegraphics*[width=9 cm]{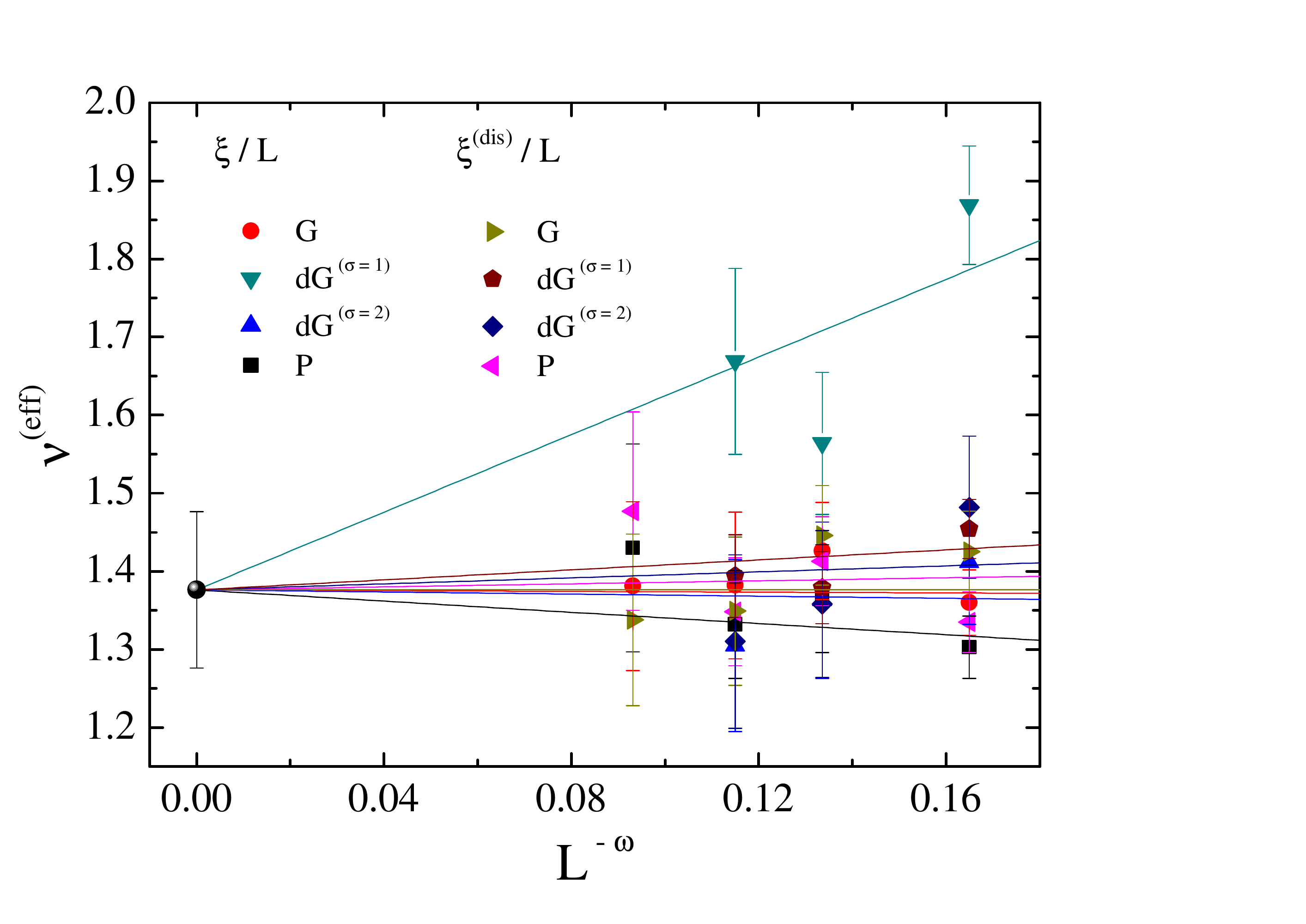}
\caption{\label{fig:nu} (Color online) Infinite limit-size
extrapolation of the effective critical exponent $\nu$.}
\end{figure}

Examples of dimensionless quantities are the connected and
disconnected correlation lengths over the system size, i.e.,
$\xi/L$ and $\xi^\mathrm{(dis)}/L$, and the Binder ratio $U_4$.
Instances of dimensionful quantities are then the derivatives of
$\xi$, $\xi^{\rm (dis)}$ ($x_\xi=1+\nu$), the connected and
disconnected susceptibilities $\chi$ and $\chi^{\rm (dis)}$
[$x_\chi= \nu(2-\eta)$,
  $x_{\chi^{\rm (dis)}}= \nu(4-\bar\eta)$], and the ratio
$U_{22}$ [$x_{U_{22}}=\nu(2\eta-\bar\eta)$] (see also
Sec.~\ref{section:observables}), which as already noted above will
serve as an alternative platform for investigating the validity of
the so-called two-exponent scaling
scenario~\cite{schwartz:85,gofman:93}.

Let us point out here that an extension of the quotients method
using the sequence of three lattice-size points $(L,2L,4L)$ will
be presented below in Sec.~\ref{section:results}. This
generalization is necessary for the scaling study of of the bond
energy of the RFIM, which is governed by a non-diverging
back-ground term.

\section{Results and Discussion}
\label{section:results}

Let us start with a few illustrations on the main heart of the
scaling method applied, that is the crossing of the universal
ratio $\xi / L$ together with the error evolution of the presented
numerical scheme. As already mentioned above, we varied $\sigma$
for the Poissonian and Gaussian distributions, see panel (a) in
Fig.~\ref{fig:crossing}, and $h_{R}$ for the double Gaussian
distribution. Panel (b) in Fig.~\ref{fig:crossing} shows the
statistical errors of the universal ratio corresponding to the
pairs of system sizes shown in panel (a) of the same figure. As
expected, the error is minimal exactly at the simulation point
denoted hereafter as $\sigma^{\rm (s)}$ or $h_{R}^{\rm (s)}$
respectively, and increases further away from it. Furthermore, a
comparative illustration with respect to the errors induced by the
reweighting method and the disorder averaging process is shown in
Fig.~\ref{fig:error_samples} again for the universal ratio $\xi /
L$ of an $L = 64$ Poissonian RFIM and for three sets of
simulations, as outlined in the figure. Clearly, this latter
accuracy test serves in favor of the proposed scheme.

\begin{figure}[htbp]
\includegraphics*[width=9 cm]{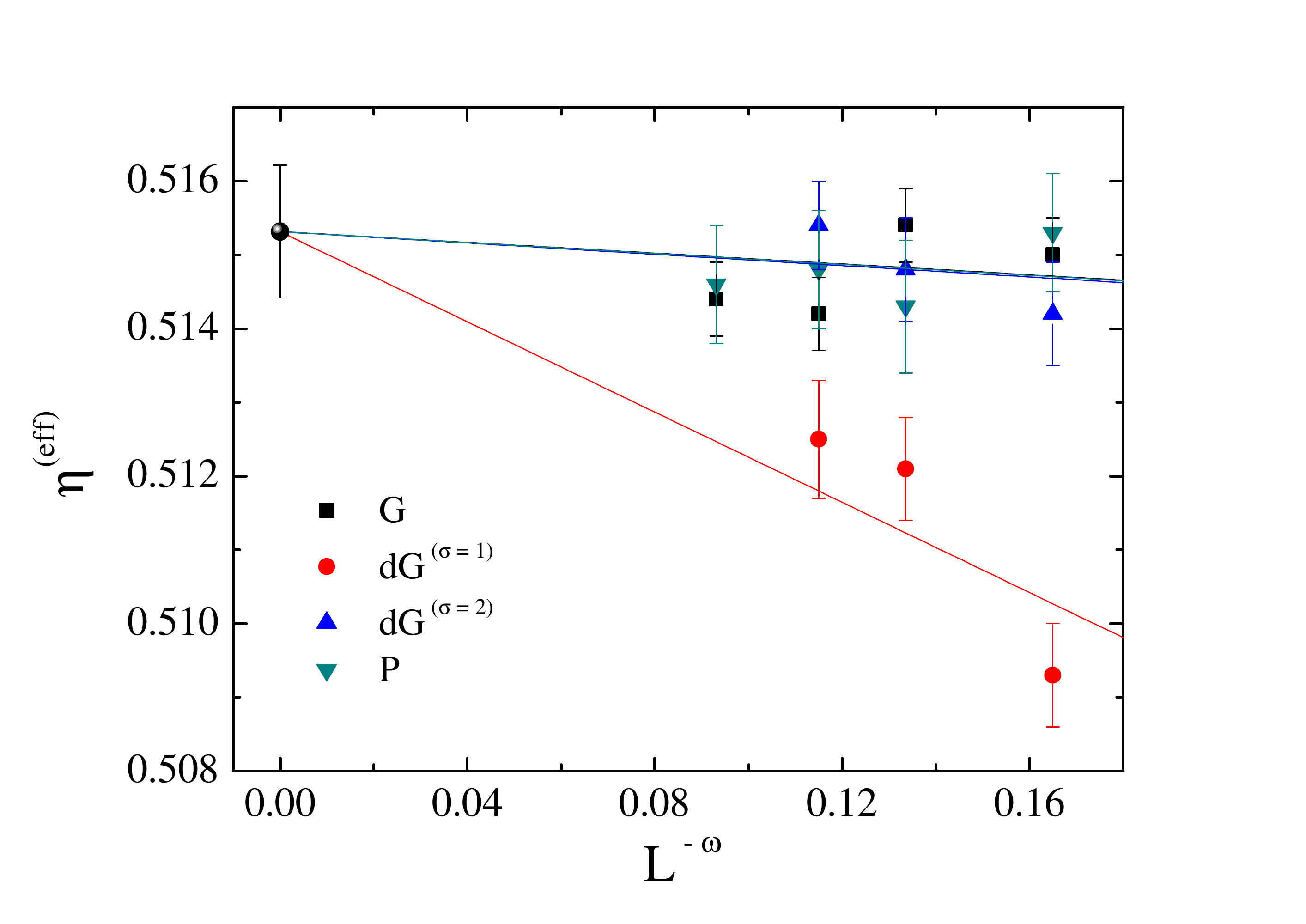}
\caption{\label{fig:eta} (Color online) Infinite limit-size
extrapolation of the effective critical exponent $\eta$. Four
solid lines are shown corresponding to the four random-field
distributions as in Fig.~\ref{fig:nu}, although they not easily
discerned due to very close values of their linear coefficient
terms.}
\end{figure}

The full application of Eq.~\eqref{eq:QO} to our four random-field
distributions has been summarized in Table II of
Ref.~\cite{fytas:13}, where all the estimates of critical points,
universal ratios, and critical exponents are given, together with
the corrections-to-scaling exponent $\omega=0.52(10)$ (see also
Fig. 4 in Ref.~\cite{fytas:13}). In particular, the computation of
the corrections-to-scaling exponent $\omega$ has been performed by
means of a joint fit, third-order polynomial in $L^{-\omega}$, for
the dimensionless quantities $\xi/L$, $\xi^{\rm (dis)}/L$, and
$U_{4}$ using data for $L\geq 24$. We should point out that joint
fits share the value of some fitting parameters such as the
$L\to\infty$ extrapolation (which is the same for all random-field
distributions), or the corrections-to-scaling exponent $\omega$
(which is common to all magnitudes). Here, we provide some
complementary illustrations, showing the infinite limit-size
extrapolation of the effective exponents of the correlation length
$\nu$, the anomalous dimension $\eta$, and the two-exponent
difference $2\eta - \bar{\eta}$, the latter serving as an
independent test of the two-exponent scaling scenario in the
theory of the random-field problem~\cite{schwartz:85}.

Figures~\ref{fig:nu} and \ref{fig:eta} illustrate the infinite
limit-size extrapolation of the effective exponents $\nu$ and
$\eta$ respectively, for all the four type of distributions
studied. The solid lines are joint polynomial fits of first order
in $L^{-\omega}$ including data points for $L\geq 32$,
extrapolating to $L^{-\omega} = 0$, as shown by the filled circle
in each figure. We remind the reader that for the effective
exponent $\nu$ we have two sets of data for each of the four
distributions coming from the connected and disconnected
correlation lengths~\cite{fytas:13}. Let us comment here that our
estimation $\nu = 1.38(10)[3]$ is similar to the most modern
computations that suggest on average a value of
$1.35(7)$~\cite{hartmann:01,middleton:01,dukovski:03,wu:05}. For
the anomalous dimension estimate $\eta = 0.5153(9)[2]$, we note
also $\eta=0.50(3)$ from Ref.~\cite{hartmann:01} as a comparison.
Obviously, our errors for $\nu$ are larger than those for $\eta$
because we compute derivatives as connected correlations
~\footnote{As in Ref.~\cite{fytas:13}, the error given in
parenthesis is of statistical origin and the one in brackets comes
from the uncertainty in the choice of $\omega$.}  (see also the
discussion in Sec.~\ref{section:FD}).

\begin{figure}[htbp]
\includegraphics*[width=9 cm]{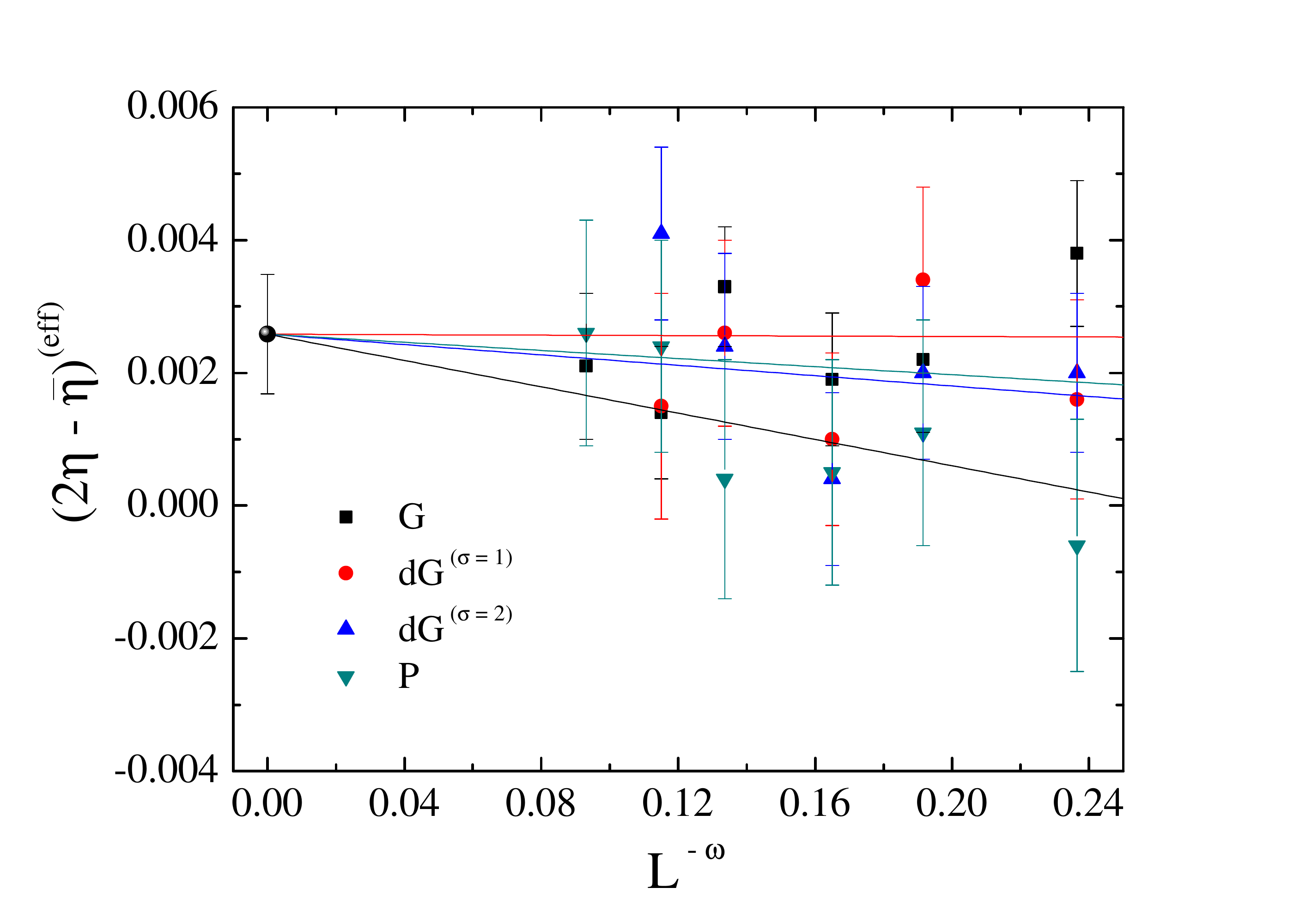}
\caption{\label{fig:U22} (Color online) Infinite limit-size
extrapolation of the effective exponent $2\eta-\bar{\eta}$ .}
\end{figure}

Subsequently in Fig.~\ref{fig:U22} we perform an extrapolation of
the effective exponent difference $2\eta - \bar{\eta}$,
corresponding to the dimensionful quantity $U_{22}$,
Eq.~\eqref{eq:U22-def}, in order to discuss the two-exponent
scaling scenario. The solid lines in this figure illustrate a
joint polynomial fit, first-order in $L^{-\omega}$, including data
points for $L\geq 16$, giving $2\eta - \bar{\eta} = 0.0026(9)[1]$
as shown by the filled black circle at $L^{-\omega} = 0$. However,
we should note here that if one fixes the infinite limit-size
point $(2\eta-\bar{\eta})|_{L=\infty}$ to zero, the fit becomes of
better quality in terms of the merit $\chi^2/{\rm
DOF}$~\cite{fytas:13}. Unfortunately, in the present $D=3$ case,
we can not draw a definite conclusion on the validity of the
two-exponent scaling scenario. Additional work is under way to
tackle this problem at higher dimensions ($D >
3$)~\cite{fytas:15}.

We turn our discussion now to the most controversial issue of the
specific heat of the RFIM. The specific heat of the RFIM can be
experimentally measured~\cite{belanger:83} and is, for sure, of
great theoretical importance. Yet, it is well known that it is one
of the most intricate thermodynamic quantities to deal with in
numerical simulations, even when it comes to pure systems. For the
RFIM, Monte Carlo methods at $T>0$ have been used to estimate the
value of its critical exponent $\alpha$, but were restricted to
rather small systems sizes and have also revealed many serious
problems, i.e., severe violations of self
averaging~\cite{parisi:02,malakis:06}. A better picture emerged
throughout the years from $T=0$ computations, proposing estimates
of $\alpha\approx 0$. However, even by using the same numerical
techniques, but different scaling approaches, some inconsistencies
have been recorded in the literature. The most prominent was that
of Ref.~\cite{hartmann:01}, where a strongly negative value of the
critical exponent $\alpha$ was estimated. On the other hand,
experiments on random field and diluted antiferromagnetic systems
suggest a clear logarithmic divergence of the specific
heat~\cite{belanger:83}.

\begin{figure}[htbp]
\includegraphics*[width=9 cm]{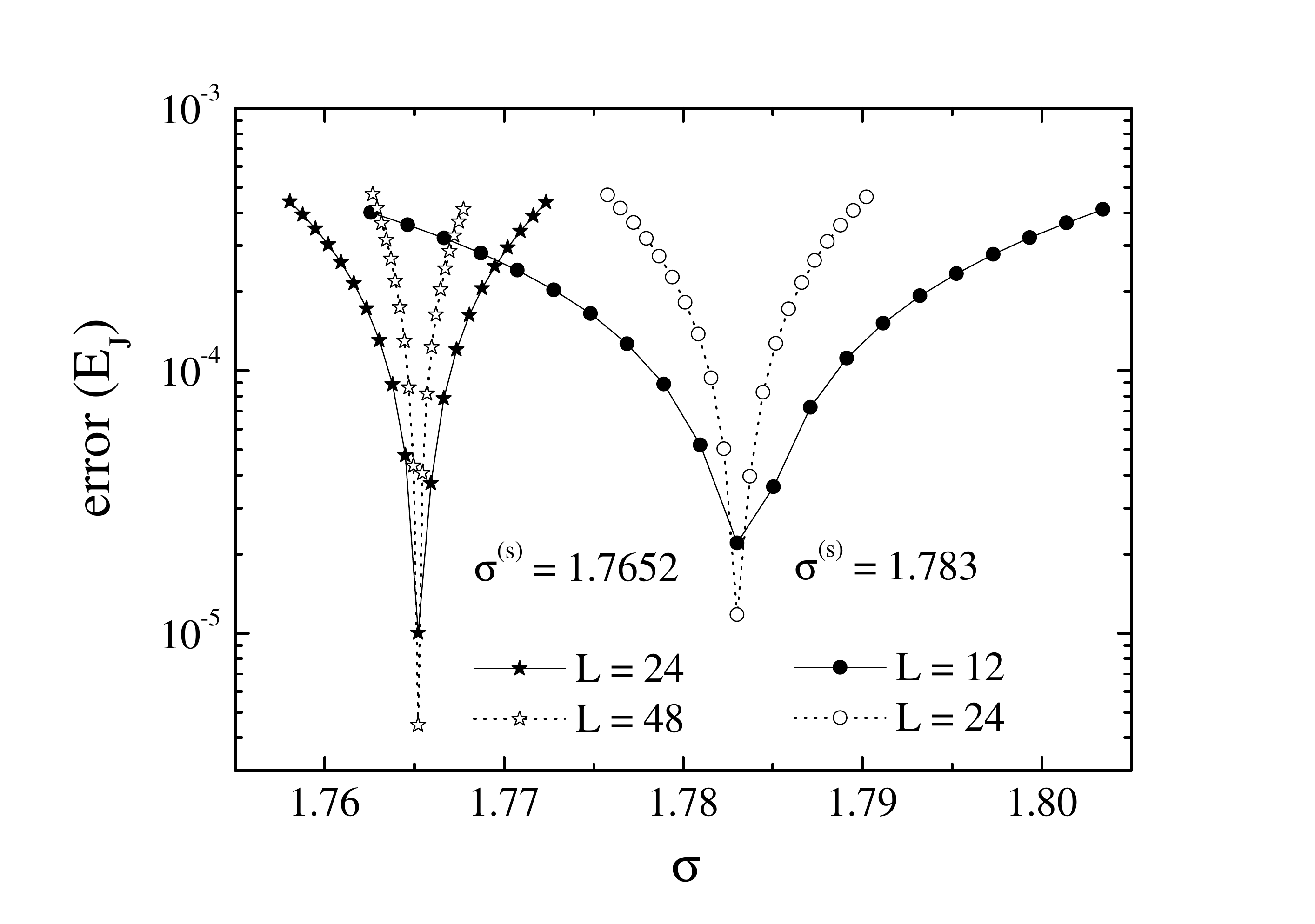}
\caption{\label{fig:error_E_J} Semi-logarithmic illustration of
statistical errors appearing in the three lattice-size variant of
the quotients method. We show results for the Poissonian RFIM and
$L_1 = 12$, $L_2 = 24$, and $L_3 = 48$. The values of the field
strength where the simulations were performed for both pairs of
system sizes are also given in the figure.}
\end{figure}

The specific heat can be estimated using ground-state calculations
and applying thermodynamic relations employed by Hartmann and
Young~\cite{hartmann:01} and Middleton and
Fisher~\cite{middleton:02}. The method relies on studying the
singularities in the bond-energy density $E_{J}$~\cite{holm:97}.
This bond energy density is the first derivative $\partial
E/\partial J$ of the ground-state energy with respect to the
random-field strength, say
$\sigma$~\cite{middleton:02,hartmann:01}. The derivative of the
sample averaged quantity $\overline{E}_{J}$ with respect to
$\sigma$ then gives the second derivative with respect to $\sigma$
of the total energy and thus the sample-averaged specific heat
$C$. The singularities in $C$ can also be studied by computing the
singular part of $\overline{E}_{J}$, as $\overline{E}_{J}$ is just
the integral of $C$ with respect to $\sigma$. The general
finite-size scaling form assumed is that the singular part of the
specific heat behaves as
\begin{equation}
\label{eq:C_scaling} C_{\rm s}\sim L^{\alpha/\nu}\tilde{C}\left
[(\sigma-\sigma^{\rm (c)})L^{1/\nu}\right].
\end{equation}
Thus, one may estimate $\alpha$ by studying the behavior of
$\overline{E}_{J}$ at $\sigma = \sigma^{\rm
(c)}$~\cite{middleton:02}. The computation from the behavior of
$\overline{E}_{J}$ is based on integrating the above scaling
equation up to $\sigma^{\rm (c)}$ , which gives a dependence of
the form
\begin{equation}
\label{eq:E_J_scaling} \overline{E}_{J}(L,\sigma = \sigma^{\rm
(c)}) = A + B L^{(\alpha-1)/\nu},
\end{equation}
with $A$ and $B$ non universal constants.

Since $\alpha-1$ is negative, Eq.~(\ref{eq:E_J_scaling}) is
dominated by the non-divergent back ground $A$, forcing us to
modify the standard phenomenological renormalization. We get rid
of $A$ by considering three lattice sizes in the following
sequence: $(L_1, L_2, L_3) = (L, 2L, 4L)$. We generalize
Eq.~(\ref{eq:QO}) by taking now the quotient of the differences
$Q_O=(O_{4L}-O_{2L})/(O_{2L} - O_{L})$ at the crossings of the
pairs $(L,2L)$ and $(2L,4L)$, respectively. Applying this formula
to the bond energy we obtain
\begin{equation}\label{eq:QO_new}
Q_{\overline{E}_J}^\mathrm{(cross)}=2^{(\alpha -
1)/\nu}+\mathcal{O}(L^{-\omega}).
\end{equation}

\begin{table}[h]
\begin{ruledtabular}
\caption{\label{tab:specheat} Effective critical exponent ratio
  $(\alpha-1)/\nu$ using a three lattice-size variant
  $(L_{1},L_{2},L_{3})=(L,2L,4L)$, see Eq.~(\ref{eq:QO_new}), of the original quotients method.}
\begin{tabular}{llc}
Distribution & $(L_{1},L_{2},L_{3})$ & $(\alpha-1)/\nu$\\
\hline
G    &$(12,24,48)$    &   -0.758(11)          \\
     &$(16,32,64)$    &   -0.793(17)         \\
     &$(24,48,96)$    &   -0.860(30)         \\
     &$(32,64,128)$   &   -0.881(75)         \\
\hline
dG$^{(\sigma=1)}$  &$(16,32,64)$    &   0.954(66)        \\
                   &$(24,48,96)$    &   -0.036(23)       \\
                   &$(32,64,128)$   &   -0.309(23)         \\
\hline
dG$^{(\sigma=2)}$  &$(12,24,48)$    &   -0.735(16)          \\
                   &$(16,32,64)$    &   -0.766(16)          \\
                   &$(24,48,96)$    &   -0.882(60)        \\
                   &$(32,64,128)$   &   -0.867(56)          \\
\hline
P   &$(12,24,48)$    &   -1.120(6)          \\
    &$(16,32,64)$    &   -1.089(10)          \\
    &$(24,48,96)$    &   -1.071(42)         \\
    &$(32,64,128)$   &   -0.970(37)         \\
\end{tabular}
\end{ruledtabular}
\end{table}

\begin{figure}[htbp]
\includegraphics*[width=9 cm]{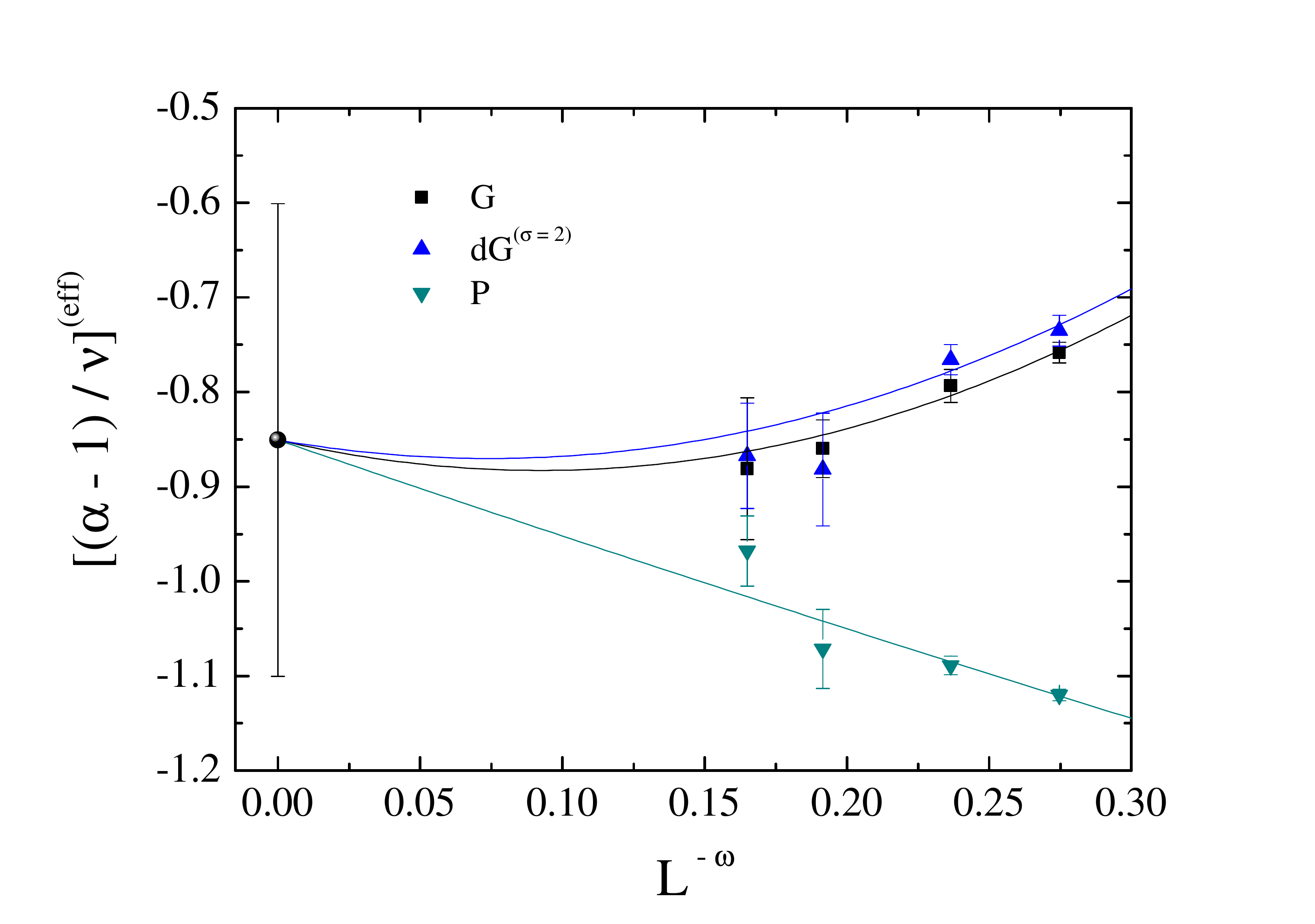}
\caption{\label{fig:alpha} (Color online) Infinite limit-size
extrapolation of the effective exponent ratio $(\alpha - 1)/\nu$.}
\end{figure}

Of course, at variance with the standard two lattice-size
phenomenological renormalization, statistical errors are
significantly amplified by the reweighting extrapolation, as it
can be clearly seen in Fig.~\ref{fig:error_E_J}. Hence, we have
preferred to carry out an independent set of simulations for
parameters corresponding to the crossing points identified in the
main analysis of the quotients method. Our results for the
effective exponent ratio $(\alpha-1)/\nu$ are given in
Table~\ref{tab:specheat} and their extrapolation is shown in
Fig.~\ref{fig:alpha}. We have excluded from the fitting procedure
the data of the double Gaussian distribution with $\sigma = 1$ as
their inclusion destabilized the fit. The solid lines in
Fig.~\ref{fig:alpha} show a joint polynomial fit, second order in
$L^{-\omega}$. The extrapolated value for the exponent ratio is
$(\alpha - 1)/\nu = -0.85(25)$ and is marked by the filled circle
in the figure at $L^{-\omega} = 0$. Using now our estimate $\nu =
1.38(10)$ for the critical exponent of the correlation length,
simple algebra (and error propagation) gives the value $\alpha =
-0.16(35)$ for the critical exponent of the specific heat. Let us
point out here that also Middleton and Fisher, using the scaling
of the bond energy at the candidate critical field value
$\sigma^{\rm (c)} = 2.27$, proposed a value of
$\alpha=-0.12(16)$~\cite{middleton:02}, compatible with our
result. Although the error proposed by the latter authors is much
smaller than ours, we have to note that their method implies an a
priori knowledge of the ``exact'' value of the critical field.
Obviously, as we have no command over this value, what is usually
done is to use some candidate values of the critical field around
the best known estimate and then repeat the simulations for all
those candidate values. However, even in this case the results are
ambiguous, as a change in the value of $\sigma^{\rm (c)}$ by a
factor of $\delta \sigma^{\rm (c)} = 10^{-3}$ results, on a
average, in a change of the order of $\delta \alpha \approx 0.04$
in the value of $\alpha$~\cite{theodorakis:13}.

Following the discussion above, our numerical studies of
disordered systems are carried out near their critical points
using finite samples; each sample is a particular random
realization of the quenched disorder. This makes it then crucial
to study the dependence of some observables with the disorder, the
so-called self-averaging properties of the system. The study of
these properties in disordered systems has generated in the past
years a large amount of
works~\cite{parisi:02,aharony:96,chamati:02,aharony:98,wiseman:98,deroulers:02,parisi:04,gordillo-guerrero:07},
still mostly focused on the bond- and site-diluted versions of the
Ising model in $D=2$ and $3$.

A typical measure of the self-averageness of a random system via a
physical quantity $A$ is given from $R_{A} = [\overline{ \langle A
\rangle^2} - \overline{\langle A \rangle}^2] / \overline{ \langle
A \rangle}^2 $. Aharony and Harris~\cite{aharony:96} predicted
that the size evolution of $R_{A}(L)$ for the random system
depends on whether the system is controlled by the pure or the
random fixed point, i.e., $R_{A}(L) \propto L^{(\alpha/\nu)_{\rm
pure}}$ for pure fixed point, and $R_{A}(L) \propto {\rm const}
\neq 0$ for random fixed point respectively, as $L\to \infty$. In
the case of the random fixed point, the system has no
self-averaging. On the other hand, the system exhibits weak
self-averaging in the case of the pure fixed point. Clearly
enough, the system is expected to be self-averaging if $R_{A} \to
0$, as $L\to \infty$.

\begin{figure}[htbp]
\includegraphics*[width=9 cm]{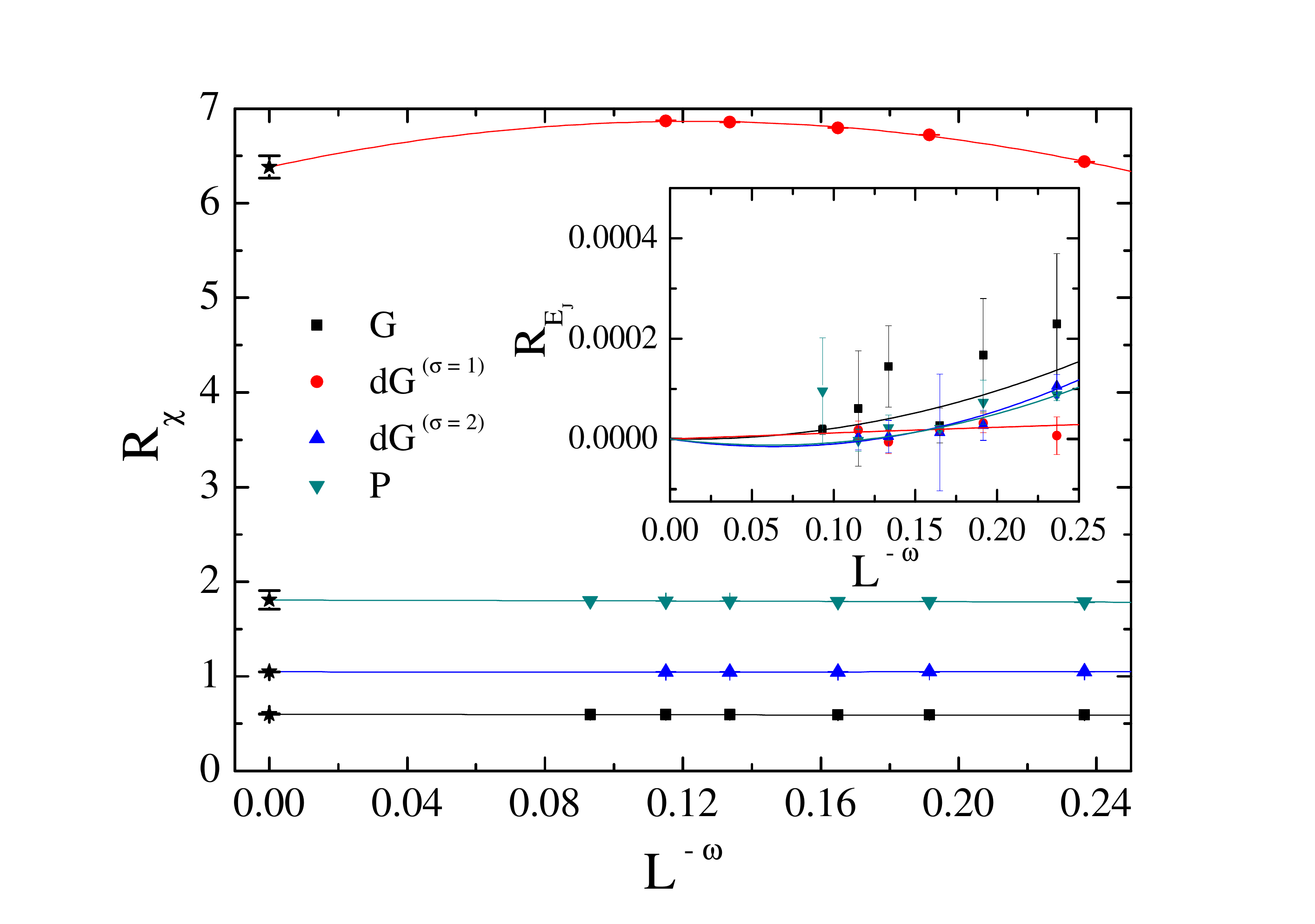}
\caption{\label{fig:self_averaging} (Color online) Infinite
limit-size extrapolation of the effective ratios $R_{\chi}$ (main
panel) and $R_{E_{J}}$ (inset).}
\end{figure}

The RFIM is a nice candidate to check the analytical predictions
of Aharony and Harris on self-averaging~\cite{aharony:96},
monitoring the infinite limit-size extrapolation of $R_{A}$. In
particular, we investigated here the behavior of the ratio for two
observables, the connected susceptibility, $R_{\chi}$, and the
bond energy, $R_{E_{J}}$. In Fig.~\ref{fig:self_averaging} we plot
the effective values of the ratios $R_{\chi}$ and $R_{E_{J}}$ in
the main panel and inset, respectively, estimated at the crossing
points of $\xi / L$ as usual, for all our four distributions
studied, as indicated by the different colors and symbols. In both
cases, the solid lines show a joint, second-order in $L^{-\omega}$
polynomial fit, using as a lower cut off the lattice size $L_{\rm
min} = 16$. For the case of $R_{\chi}$, the extrapolated values
shown by black stars, are dependent on the field distribution and
are clearly non-zero, indicating violation of self-averaging.
In particular, we quote the following limiting values: $R_{\chi} =
\{0.600(2), \;6.38(11), \; 1.047(4), \; 1.81(9)\}$, for the cases
of the Gaussian, double Gaussian with $\sigma=1$, double Gaussian
with $\sigma=2$, and Poissonian distributions, respectively. The
above results verify the prediction of Ref.~\cite{aharony:96},
according to which the susceptibility at the critical point is not
self-averaging for models where the disorder is relevant, relevant
meaning that the critical exponent of the specific heat of the
corresponding pure model is positive ($\alpha_\mathrm{pure}>0$)
\footnote{Note, however, that we do not compute the susceptibility
for each sample.  Rather, we compute quantities $\hat\chi$
tailored in such a way that their average over the random fields
is the susceptibility $\chi$, recall
Eqs.~(\ref{eq:h_k_G}),~(\ref{eq:h_k_P}), and (\ref{eq:h_k_d_G}).
The self-averageness ratio is computed from such $\hat\chi$.}. As
for the self-averaging ratio for the bond energy, shown in
Fig.~\ref{fig:self_averaging}--inset, it goes to zero with
increasing system size, indicating self-averaging in the
thermodynamic limit.

Finally, we present some computational aspects of the implemented
push-relabel algorithm and its performance on the study of the
RFIM. Although its generic implementation has a polynomial time
bound, its actual performance depends on the order in which
operations are performed and which heuristics are used to maintain
auxiliary fields for the algorithm. Even within this polynomial
time bound, there is a power-law critical slowing down of the
push-relabel algorithm at the zero-temperature
transition~\cite{ogielski:86,middleton:02b}. This critical slowing
down is certainly reminiscent of the slowing down seen in local
algorithms of statistical mechanics at finite temperature, such as
the Metropolis algorithm, and even for cluster algorithms. In
fact, Ogielski~\cite{ogielski:86} was the first to note that the
push-relabel algorithm takes more time to find the ground state
near the transition in three dimensions from the ferromagnetic to
paramagnetic phase.

\begin{figure}[htbp]
\includegraphics*[width=9 cm]{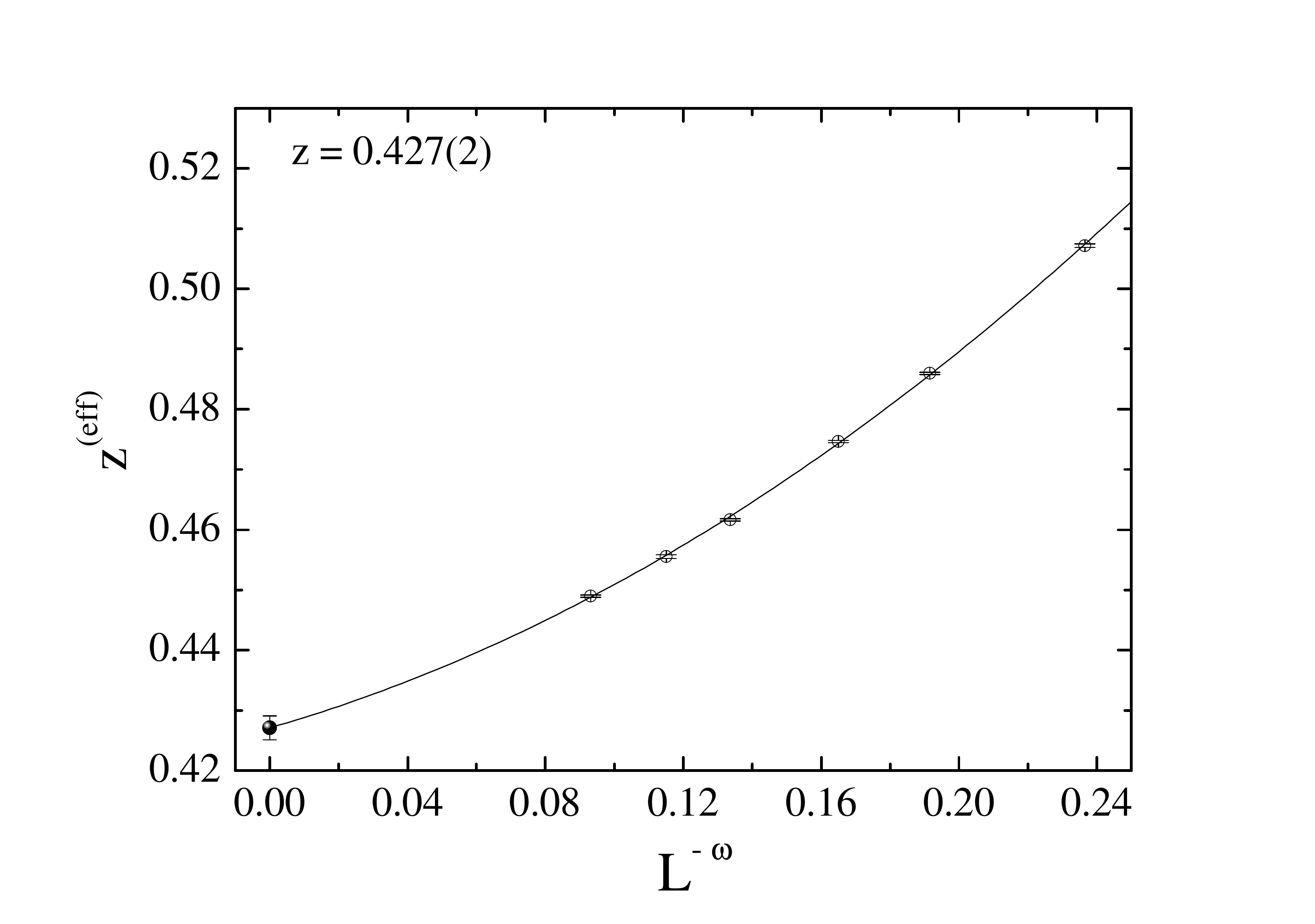}
\caption{\label{fig:exponent_z} Infinite limit-size extrapolation
of the effective exponent $z$ of the push-relabel algorithm.}
\end{figure}

A direct way to measure the dynamics of the algorithm is to
examine the dependence of the running time, measured by the number
of push-relabel operations, on system size $L$. Such an analysis
has already been performed in Ref.~\cite{meinke:05} for the
Gaussian $D=3$ RFIM and a FIFO queue implementation, as in the
current paper, finding a dynamic exponent $z = 0.43(6)$, using the
data collapse technique and fixing the values $\sigma^{\rm
(c)}=2.27$ and $\nu=1.37$ in the scaling ansatz. Here, we present
a complementary analysis based on the numerical data also for
Gaussian RFIM, using our scaling approach within the quotients
method and without assuming prior knowledge of the critical field
and correlation length exponent. Our fitting attempt is shown in
Fig.~\ref{fig:exponent_z}, where the solid line is a second-order
in $L^{-\omega}$ polynomial for system sizes $L\ge 16$ and the
obtained estimate for the dynamic critical exponent $z$ is
$0.427(2)$, very close to the estimate of Ref.~\cite{meinke:05}.

\section{Summary and outlook}
\label{section:summary}

To summarize, we have presented in the current paper a new
approach to the study of the random-field Ising model, using as a
platform the three-dimensional version of the model. We combined
several efficient numerical methods, from zero-temperature
optimization algorithms to generalized fluctuation-dissipation
formulas and reweighting extrapolations that allowed the
computation of response functions, as well as advanced finite-size
scaling techniques that offered us the possibility to tackle some
of the hardest open problems in the random-field literature, like
the existence and role of scaling corrections and the universality
principle of the model. We hope that this contribution gives a
clear overview of all the technical details of our implementation,
paving the way to even more sophisticated studies in the field of
disordered systems. Currently, using the prescription outlined
above, we are dealing with the random-field problem at higher
dimensions and we expect to provide clear-cut results regarding
the validity of the two-exponent scaling scenario, one of the
building blocks in the scaling theory of the random-field Ising
model.

\begin{acknowledgments}
We are grateful to D. Yllanes and, especially, to L.A.
Fern\'{a}ndez for substantial help during several parts of this
work. We also thank M. Picco and N. Sourlas for reading the
manuscript. We were partly supported by MINECO, Spain, through
research contract No. FIS2012-35719-C02-01. Significant
allocations of computing time were obtained in the clusters
\emph{Terminus} and \emph{Memento} (BIFI). N.G. Fytas acknowledges
financial support from a Royal Society Research Grant under No
RG140201 and from a Research Collaboration Fellowship Scheme of
Coventry University.
\end{acknowledgments}

\bibliographystyle{apsrev4-1}
\bibliography{biblio.bib,biblio_nikos.bib}

\end{document}